\documentclass[aps,prd,reprint,groupeaddress,showpacs,preprintnumbers,nofootinbib,nobibnotes,amsmath,amssymb,floatfix,superscriptaddress,groupedaddress]{revtex4-1}
\usepackage[colorlinks=true,linkcolor=red,citecolor=blue,urlcolor=blue]{hyperref}
\usepackage{amssymb}
\usepackage{epsfig} 
\usepackage{amsmath} 
\usepackage{graphicx}
\usepackage{color}
\usepackage{float}
\usepackage{xcolor}
\usepackage{amssymb}
\usepackage{enumitem}
\usepackage{natbib}
\usepackage[T1]{fontenc}
\definecolor{red}{rgb}{0.8,0,0}
\definecolor{violet}{rgb}{0.4,0,0.4}
\definecolor{green}{rgb}{0,0.5,0.0}
\definecolor{navy}{rgb}{0.0,0.0,0.6}
\definecolor{orange}{rgb}{0.8,0.2,0.0}
\usepackage[normalem]{ulem}  % \sout{old text} for strikeout

\newcommand{\physrep}{Phys. Rep.}
\newcommand{\apjl}{Astro. Phys. J. Lett.}
\newcommand{\mnras}{Mon. Not. Roy. Astron. Soc.}
\newcommand{\aap}{Astronomy $\&$ Astrophysics}
\newcommand{\nphysa}{Nuc. Phys. A}

\begin{document} 

\title{Influence of the nuclear symmetry energy slope on observables of
compact stars with $\Delta$-admixed hypernuclear matter}

\author{Vivek Baruah Thapa}
\email{thapa.1@iitj.ac.in}
\author{Monika Sinha}
\email{ms@iitj.ac.in}
\affiliation{Indian Institute of Technology Jodhpur, Jodhpur 342037, India}

\date{\today}
\begin{abstract}
{In this work, we study the effects of nuclear symmetry energy slope on neutron star dense matter equation of state and its impact on neutron star observables (mass-radius, tidal response).
We construct the equation of state within the framework of covariant density functional theory implementing coupling schemes of non-linear and density-dependent models with viability of heavier non-nucleonic degrees of freedom.
The slope of symmetry energy parameter ($L_{\text{sym}}$) is adjusted following density-dependence of isovector meson coupling to baryons.
We find that smaller values of $L_{\text{sym}}$ at saturation favour early appearance of $\Delta$-resonances in comparison to hyperons leading to latter's threshold at higher matter densities. We also investigate the dependence of $L_{\text{sym}}$ on tidal deformability and compactness parameter of a $1.4~M_\odot$ neutron star for different equation of states and observe similar converging behaviour for larger $L_{\text{sym}}$ values.}
\end{abstract}

\maketitle

\section{Introduction}\label{sec:intro}

In stellar evolution, when nuclear fusion stops and the electron degeneracy pressure can not prevent the gravitational collapse, electrons happen to combine with protons and produce neutron rich nuclei. Eventually neutrons drip out of the nuclei, and the matter becomes mainly composed of neutrons with some admixture of protons and electrons. This remnant of stellar evolution is called neutron star (NS). Thus, after the type-II supernova explosion the stellar remnant forms highly dense NS. Naturally, the matter density inside the NS varies from subnuclear density near surface to supranuclear density towards the centre. Matter is composed of ions and electrons near the surface, i.e. in the outer crust region, and in the inner crust region neutron-rich nuclei and some free neutrons appear. Then in the core, matter is completely made of free neutrons with comparatively less number of protons and electrons. With the increase of density towards the centre of the star, exotic components of matter viz. hyperons, heavier non-strange baryons, Boson condensates, even deconfined quarks may appear while near surface where density is comparatively low up to $\approx 2$ times nuclear saturation density ($n_0$), the matter is composed of only nucleons \cite{1996cost.book.....G, weber2017pulsars}. Hence, the interior of NS is a good domain to study the dense nuclear matter in bulk with and without exotic degrees of freedom. In general, we have good knowledge about finite nuclei at $n_0$. Hence, the theoretical idea of uniform symmetric nuclear matter in bulk even at varied densities above $n_0$ is just the extrapolation and idealization of finite nuclei knowledge. On the other hand, the observational test of bulk matter properties in varied density above $n_0$ can only be obtained from astrophysical observations of compact NSs.

One important ingredient for energy density of nucleons inside nuclear matter is the symmetry energy ($E_{\text{sym}}$). Consequently, nuclear symmetry energy and its density dependence play a salient role in comprehending dense matter behaviour \cite{2010EL.....9132001D, 2016PhR...621..127L}. In case of finite nuclei, the contribution of symmetry energy to the mass of the nuclei is small compared to other terms in the semi-empirical mass formula. Hence, the experimental information regarding $E_{\text{sym}}$ is not very sound even at $n_0$. As mentioned earlier, from the astrophysical properties of NSs the nuclear matter properties at varied density can be determined. The nuclear symmetry energy and its variation with density affects substantially the composition and matter pressure what in consequence affects the NS properties specially the radius \cite{PhysRevLett.102.122502, PhysRevLett.120.172702}. Although from nuclear data we have very little knowledge of $E_{\text{sym}}$, a good comprehensive idea of the same can now be drawn from the recent radius measurement of certain NS candidates PSR J$0030+0451$ \cite{2019ApJ...887L..24M,2019ApJ...887L..21R, 2020PhRvD.101l3007L} and  PSR J$0740+6620$ \cite{2021arXiv210506980R, 2021arXiv210506979M} from NICER (Neutron star Interior Composition ExplorER) space mission. The detection of gravitational wave (GW) emissions from binary NS mergers (GW170817 \cite{LIGO_Virgo2017c, LIGO_Virgo2018a, PhysRevX.9.011001}, GW190425 \cite{2020ApJ...892L...3A}) by the LIGO-Virgo Collaboration (LVC) marked an appreciable breakthrough in the domain of multi-messenger astronomy. The GW observations set bound on the mutual tidal deformability ($\tilde{\Lambda}$) which also depends on the matter property linked with $E_{\text{sym}}$.

Not only that, comprehensive idea regarding symmetry energy behaviour can be gathered via studying its effects on other NS properties such as maximum mass, compactness, tidal deformability.
The observational estimations of certain parameters viz. maximum mass and radius of compact objects impose vital constraints on narrowing down to a unique dense matter equation of state (EOS). For instance, with the observations of massive compact stars ($M\geq 2~M_{\odot}$) such as PSR J$1614-2230$ ($1.908\pm 0.016~M_\odot$) \cite{2010Natur.467.1081D, Arzoumanian_2018}, PSR J$0348+0432$ ($2.01\pm 0.04~M_\odot$) \cite{2013Sci...340..448A}, PSR J$0740+6620$ ($2.08^{+0.07}_{-0.07}$ $M_\odot$ with 68.3\% credibility \cite{2021arXiv210400880F}), the soft EOSs tend to be invalid. The mass-radius measurements of PSR J$0030+0451$ \cite{2019ApJ...887L..24M,2019ApJ...887L..21R, 2020PhRvD.101l3007L} and PSR J$0740+6620$ \cite{2021arXiv210506980R, 2021arXiv210506979M} from NICER space mission provide significant constraint on dense matter EOS.
The latest NICER mass-radius measurement (PSR J$0740+6620$) suggests repulsive matter behaviour at higher density regimes. Joint analysis of data from NICER observations for PSR J$0030+0451$ and GW170817 event provides bounds on NS properties viz. tidal deformability ($\Lambda_{1.4}$) and radius ($R_{1.4}$) for a $1.4 M_\odot$ star \cite{Jiang_2020}.

Many recent studies have been done to constrain the values of $E_{\text{sym}}$ at $n_0$ and its slope ($L_{\text{sym}}$) at $n_0$ based on data from various astrophysical observations as well as terrestrial experiments \cite{2014EPJA...50...40L, 2015PhRvC..92f4304R, 2017ApJ...848..105T}.
Very recently an improved value of neutron skin thickness of $^{208}\text{Pb}$ was reported in Lead Radius EXperiment-II (PREX-2) to be $R_{\text{skin}}=R_n - R_p=(0.283\pm 0.071)$ fm \cite{PhysRevLett.126.172502}. This evaluates the corresponding symmetry energy and its slope to be $E_{\text{sym}}=(38.1 \pm 4.7)$ MeV and $L_{\text{sym}}=(106\pm 37)$ MeV respectively at $n_0$ with correlation coefficient as 0.978 \cite{PhysRevLett.126.172503}.
These updated values of isospin asymmetry parameters are larger than the ones ($28.5$ MeV $\leq E_{\text{sym}} (n_0) \leq 34.9$ MeV; $30.6$ MeV $\leq L_{\text{sym}} (n_0) \leq 86.8$ MeV) previously reported in ref.-\cite{2017RvMP...89a5007O} obtained by comparison of experimental data from finite nuclei and heavy-ion collisions with different microscopic model calculations. 
In this work, we explore the influence of nuclear symmetry energy on dense matter EOS, consequently on NS properties. As a consequence, we attempt to constrain the $E_{\text{sym}}$ and $L_{\text{sym}}$ from the observational features of NSs. 
To do so, we consider covariant density functional (CDF) model implementing the density-dependence of isovector-vector coupling as introduced in ref.-\cite{2017PhDT.......135S} and incorporating non-linear GM1 \cite{1991PhRvL..67.2414G} and density dependent DD-MEX \cite{TANINAH2020135065} coupling parametrizations. In the CDF scheme, the coupling constants are chosen in such a way that the model can reproduce the experimental quantities known at $n_0$. Thus the observational properties of NSs which depends on the EOS parameters {\it i.e.} coupling constants will determine $E_{\text{sym}}$ and $L_{\text{sym}}$. In several previous studies 
the symmetry energy effects on dense matter has been considered with the matter composition to be purely nucleonic \cite{2019PhRvC.100d5801J, 2020PTEP.2020d3D01H, 2021PhRvC.104a5802W}. However, due to increasing Fermi energy of nucleons interior to NS, the appearance of additional degrees of freedom such as hyperons \cite{1991PhRvL..67.2414G, Weissenborn2012a, Bonanno2012A&A, Colucci_PRC_2013, Oertel2015, Tolos2017b,  2018MNRAS.475.4347R,2018EPJA...54..133L,2021NuPhA100922171L, 2021arXiv210702245L}, $\Delta$-resonances \cite{Drago_PRC_2014, Cai_PRC_2015, Zhu_PRC_2016, Sahoo_PRC_2018, Kolomeitsev_NPA_2017, Li_PLB_2018, Li2019ApJ, Ribes_2019, Li2020PhRvD, particles3040043, 2021MNRAS.507.2991T}, meson ($\pi,\bar{K},\rho$) condensations \cite{particles2030025, 1982ApJ...258..306H, 1999PhRvC..60b5803G, 2001PhRvC..63c5802B, 1997PhR...280....1P, 2021EPJST.tmp...32M, PhysRevD.102.123007, PhysRevD.103.063004} are inevitable or energetically favorable for massive stars as recently observed \cite{2010Natur.467.1081D, Arzoumanian_2018, 2013Sci...340..448A, 2021arXiv210400880F}.
Moreover, the appearance of these exotic matter softens the EOS reducing maximum mass of the stars which is in contrary to the observations. Hence, the choice of EOS parametrization is limited with the observational constraints of massive stars. For that we have chosen two EOS parametrizations within CDF model viz. GM1 with nonlinear model and DD-MEX with density dependent coupling constant which satisfy the observational constraints of massive stars. This work will therefore explore the novel aspects of density-dependent isovector coupling on dense matter EOS with the onset of heavier strange and non-strange degrees of freedom and study the symmetry energy slope effects on NS properties.

The paper is organized as follows.
In Sec.-\ref{sec:formalism}, we briefly describe the CDF model formalism, its extension to additional heavier degrees of freedom and coupling parameters for constructing the EOS.
The effects of nuclear symmetry energy on dense matter are shown and discussed in Sec.-\ref{sec:results}.
Sec.\ref{sec:summary} provides the summary and conclusions of this work.

\textit{Conventions}: We implement the natural units $G=\hbar=c=1$ throughout the work.

\section{Formalism}\label{sec:formalism}

\subsection{CDF Model}

This section briefly describes the CDF model implemented in this work to construct the dense matter EOS.
The dense matter composition considered in this work are the entire baryon octet ($N\equiv n,p$; $Y\equiv \Lambda^0,\Sigma^{\pm,0}, \Xi^{-,0}$) and $\Delta$-resonances ($\Delta \equiv \Delta^+, \Delta^{++}, \Delta^0, \Delta^-$) along with leptons ($l\equiv e^-,\mu^-$).
In order to mediate the effective interactions between the baryons, isoscalar-scalar $\sigma$, isoscalar-vector $\omega$ and isovector-vector $\rho$-mesons are considered.
An additional hidden strangeness isoscalar-vector $\phi$-meson is also brought into consideration to describe the hyperon-hyperon repulsive interactions.
The total Lagrangian density is given by \cite{1996cost.book.....G,Li_PLB_2018}
\begin{equation}\label{eqn.1}
\begin{aligned}
\mathcal{L} & = \sum_{b\equiv N,Y} \bar{\psi}_b(i\gamma_{\mu} D^{\mu}_{(b)} - m^{*}_b) \psi_b + \sum_{l} \bar{\psi}_l (i\gamma_{\mu} \partial^{\mu} - m_l)\psi_l 
\\
& + \sum_{\Delta} \bar{\psi}_{\Delta \nu}(i\gamma_{\mu} D^{\mu}_{(\Delta)} - m^{*}_{\Delta}) \psi^{\nu}_{\Delta} + \frac{1}{2}(\partial_{\mu}\sigma\partial^{\mu}\sigma - m_{\sigma}^2 \sigma^2) \\
 & -  \frac{1}{4}\omega_{\mu\nu}\omega^{\mu\nu} + \frac{1}{2}m_{\omega}^2\omega_{\mu}\omega^{\mu} - \frac{1}{4}\boldsymbol{\rho}_{\mu\nu} \cdot \boldsymbol{\rho}^{\mu\nu} + \frac{1}{2}m_{\rho}^2\boldsymbol{\rho}_{\mu} \cdot \boldsymbol{\rho}^{\mu}  \\
 & - \frac{1}{4}\phi_{\mu\nu}\phi^{\mu\nu} + \frac{1}{2}m_{\phi}^2\phi_{\mu}\phi^{\mu} - \text{U}(\sigma),
\end{aligned}
\end{equation}
where the baryon octet, lepton Dirac and $\Delta$ baryon Schwinger-Rarita fields are represented by $\psi_b, \psi_l$ and $\psi_\Delta$ respectively.
The covariant derivative is given by $D_{\mu (j)} = \partial_\mu + ig_{\omega j} \omega_\mu + ig_{\rho j} \boldsymbol{\tau}_{j3} \cdot \boldsymbol{\rho}_{\mu} + ig_{\phi j} \phi_\mu$ with $j$ denoting the baryon particle spectrum.
The isospin projection of the third component of isovector-vector meson field is represented by $\boldsymbol{\tau}_{j3}$.
The scalar self-interaction term, which is present only in non-linear model, introduced to account for the incompressibility \cite{1996cost.book.....G} is given by
\begin{equation} \label{eqn.2}
\text{U}(\sigma)=\frac{1}{3}g_2\sigma^3 + \frac{1}{4}g_3\sigma^4
\end{equation}
where $g_2$, $g_3$ are the coefficients of self-interactions. 
$\omega_{\mu \nu}$, $\boldsymbol{\rho}_{\mu \nu}$ and $\phi_{\mu \nu}$ are the anti-symmetric field tensors corresponding to vector meson fields and given by
 \begin{equation} \label{eqn.3}
  % -----------------------------------------------------
  \begin{aligned}
  \omega_{\mu \nu} & = \partial_{\mu}\omega_{\nu} - \partial_{\nu}\omega_{\mu} ,\\
  \boldsymbol{\rho}_{\mu \nu} & = \partial_{\mu}
  \boldsymbol{\rho}_{\nu} - \partial_{\nu}\boldsymbol{\rho}_{\mu}, \\
  \phi_{\mu \nu} & = \partial_{\mu}\phi_{\nu} - \partial_{\nu}\phi_{\mu} .
\end{aligned}
\end{equation}
The Dirac and Schwinger-Rarita effective masses are respectively given by
\begin{equation} \label{eqn.4}
\begin{aligned}
    m_{b}^* & = m_b - g_{\sigma b}\sigma,\quad
    m_{\Delta}^* = m_\Delta - g_{\sigma \Delta}\sigma
  \end{aligned}
\end{equation}
with $m_b$, $m_{\Delta}$ representing the bare masses of baryon octet and $\Delta$-quartet particle spectrum respectively.
In mean-field approximation, the non-vanishing meson fields obtained by solving the Euler-Lagrange equations are given by
\begin{equation}\label{eqn.5}
\begin{aligned}
\sigma & = -\frac{1}{m_{\sigma}^2} \frac{\partial \text{U}}{\partial \sigma} + \sum_{b} \frac{1}{m_{\sigma}^2} g_{\sigma b}n_{b}^s + \sum_{\Delta} \frac{1}{m_{\sigma}^2} g_{\sigma \Delta}n_{\Delta}^s,\\
  \omega_{0} & = \sum_{b} \frac{1}{m_{\omega}^2} g_{\omega b}n_{b} + \sum_{\Delta} \frac{1}{m_{\omega}^2} g_{\omega \Delta}n_{\Delta}, \\
    \rho_{03} & = \sum_{b} \frac{1}{m_{\rho}^2} g_{\rho b}
  \boldsymbol{\tau}_{b3}n_{b} + \sum_{\Delta} \frac{1}{m_{\rho}^2} g_{\rho \Delta}
  \boldsymbol{\tau}_{\Delta 3}n_{\Delta}, \\
     \phi_{0} & = \sum_{Y} \frac{1}{m_{\phi}^2} g_{\phi Y}n_{Y}.
\end{aligned}
\end{equation}
The scalar and vector (number) densities of the constituent particle spectrum are given by $n^s = \langle\bar{\psi} \psi \rangle$ and $n=\langle\bar{\psi} \gamma^0 \psi \rangle$ respectively.
The chemical equilibrium conditions between species in the particle spectrum when strangeness is not conserved is given by \cite{2001PhRvC..63c5802B, Drago_PRC_2014}
\begin{equation} \label{eqn.6}
\mu_j = \mu_n - q_j \mu_e
\end{equation}
where $q_j$ is the charge of $j$th baryon, $\mu_e$, $\mu_n$ denote the chemical potential of electron and neutron respectively.
The chemical potential of $j$th baryon is defined as
\begin{equation}\label{eqn.7}
\begin{aligned}
	& \mu_{j} = \sqrt{p_{F_j}^2 + m_{j}^{*2}} + \Sigma^{0} + \Sigma^{r},
\end{aligned}
\end{equation}
with $\Sigma^{0} = g_{\omega j}\omega_{0} + g_{\phi j}\phi_{0} + g_{\rho j} \boldsymbol{\tau}_{j3} \rho_{03}$ and the re-arrangement term necessary to maintain thermodynamic consistency is given by
\begin{equation}
\begin{aligned}\label{eqn.8}
	\Sigma^{r} & = \sum_{b} \left[ \frac{\partial g_{\omega b}}{\partial n}\omega_{0}n_{b} - \frac{\partial g_{\sigma b}}{\partial n} \sigma n_{b}^s + \frac{\partial g_{\rho b}}{\partial n} \rho_{03} \boldsymbol{\tau}_{b3} n_{b} \right. \\
	& \left. + \frac{\partial g_{\phi b}}{\partial n}\phi_{0}n_{b} \right] + \sum_{\Delta} (\psi_b \longrightarrow \psi_{\Delta}^{\nu}).
\end{aligned}
\end{equation}
In case of non-linear coupling model for this present work, only the isovector $\rho$-meson coupling contributes to re-arrangement term.

The dense matter EOS is calculated self-consistently taking into account two additional constraints, viz. charge neutrality and global baryon number conservation respectively given by
\begin{equation} \label{eqn.9}
\begin{aligned}
& \sum_b q_b n_b + \sum_\Delta q_\Delta n_\Delta - n_e - n_\mu = 0, \\
& \sum_b n_b + \sum_\Delta n_\Delta = n.
\end{aligned}
\end{equation}
The total energy density is given by
\begin{eqnarray} \label{eqn.10}
\begin{aligned}
\varepsilon & = \frac{1}{2}m_{\sigma}^2 \sigma^{2} + \frac{1}{2} m_{\omega}^2 \omega_{0}^2 + \frac{1}{2}m_{\rho}^2 \rho_{03}^2 + \frac{1}{2} m_{\phi}^2 \phi_{0}^2 \\
& + \sum_{j\equiv b,\Delta} \frac{2J_j + 1}{2 \pi^2} \left[ p_{{F}_j} E^3_{F_j} - \frac{m_{j}^{*2}}{8} \left( p_{{F}_j} E_{F_j} \right. \right. \\ 
& \left. \left. + m_{j}^{*2} \ln \left( \frac{p_{{F}_j} + E_{F_j}}{m_{j}^{*}} \right) \right) \right] + \frac{1}{\pi^2}\sum_l \left[ p_{{F}_l} E^3_{F_l} \right. \\
& \left. - \frac{m_{l}^{2}}{8} \left( p_{{F}_l} E_{F_l} + m_{l}^{2} \ln \left( \frac{p_{{F}_l} + E_{F_l}}{m_{l}} \right) \right) \right].
\end{aligned}
\end{eqnarray}
The matter pressure is then evaluated following the Gibbs-Duhem relation defined as
\begin{equation} \label{eqn.11}
P = \sum_{j\equiv b,\Delta} \mu_j n_j + \sum_{l} \mu_l n_l - \varepsilon.
\end{equation}

\subsection{NS structure and properties}

The NS properties (mass-radius) are deduced from the Tolman-Oppenheimer-Volkoff (TOV) equations for non-rotating, spherically symmetric NS configurations corresponding to constructed EOSs which are given by \cite{1996cost.book.....G}
\begin{equation} \label{eqn.12}
\begin{aligned}
\frac{dP(r)}{dr} & =-\frac{[\varepsilon(r)+P(r)][M(r)+4\pi r^3 P(r)]}{r^2[1-2M(r)/r]}, \\
\frac{dM(r)}{dr} & =4\pi r^2 \varepsilon(r),
\end{aligned}
\end{equation}
where, $M(r)$ is the gravitational mass enclosed within radius $r$.
Solutions of TOV equations are obtained on applying the boundary conditions, $P(R)=M(0)=0$.

\begin{table*} %[h!]
\centering
\caption{(i) Parameter values of CDF coupling models considered in this work.}
\begin{tabular}{cccccccccc}
\hline \hline
Coupling & \multicolumn{1}{c}{$g_{\sigma N}$} & $g_{\omega N}$ & $g_{\rho N}$ & $g_2$ & $g_3$ & $m_\sigma$ & $m_\omega$ & $m_\rho$ & $m_\phi$ \\
Model & & & & (fm$^{-1}$) & & (MeV) & (MeV) & (MeV) & (MeV) \\
\hline
GM1 & 9.5708 & 10.5964 & 8.1957 & 12.2817 & $-8.9780$ & 550 & 783 & 770 & 1019.45 \\
DD-MEX & 10.7067 & 13.3388 & 7.2380 & $-$ & $-$ & 547.3327 & 783 & 763 & 1019.45 \\
\hline \hline
\multicolumn{10}{c}{(ii) Coefficient values corresponding to eqns.-\eqref{eqn.18}, \eqref{eqn.func} in DD-MEX coupling model.} \\
\hline
\multicolumn{2}{c}{Meson ($i$)} & \multicolumn{2}{c}{$a_i$} & \multicolumn{2}{c}{$b_i$} & \multicolumn{2}{c}{$c_i$} & \multicolumn{1}{c}{$d_i$} &  \\
\hline
\multicolumn{2}{c}{$\sigma$} & \multicolumn{2}{c}{1.3970} & \multicolumn{2}{c}{1.3350} & \multicolumn{2}{c}{2.0671} & \multicolumn{1}{c}{0.4016} & \\
\multicolumn{2}{c}{$\omega$} & \multicolumn{2}{c}{1.3926} & \multicolumn{2}{c}{1.0191} & \multicolumn{2}{c}{1.6060} & \multicolumn{1}{c}{0.4556} & \\
\multicolumn{2}{c}{$\rho$} & \multicolumn{2}{c}{0.6202} & \multicolumn{2}{c}{} & \multicolumn{2}{c}{} & \multicolumn{1}{c}{} & \\
\hline
\end{tabular}
\label{tab:1}
\end{table*}

The tidal response of compact stars to an external gravitational field is quantified in terms of dimesionless tidal deformability defined as \cite{Hinderer_2008}
\begin{equation}\label{eqn.13}
\Lambda=\frac{2}{3} k_2 \left(\frac{M}{R}\right)^{-5}  
\end{equation}
where $k_2$ is the tidal Love number defined as
\begin{equation}
\begin{aligned}
k_2 = & \frac{8 C^5}{5}(1-2C^2)[2+2C(y-1)-y] \cdot \\
& \{2C [6-3y+3C(5y-8)] + 4C^3 [13 - 11y \\
& + C(3y-2) + 2C^2(1+y)] + 3(1-2C^2) \cdot \\
& [2 - y + 2C(y-1)] \log (1-2C)\}^{-1}
\end{aligned}
\end{equation}
and obtained by solving the differential equation \cite{PhysRevD.80.084018},
\begin{equation}\label{eqn.14}
r \frac{dy(r)}{dr} + y(r)^2 + y(r)F(r) + r^2 Q(r) = 0,
\end{equation}
with boundary condition as $y(0)=2$ where the functions are defined as,
\begin{equation}\label{eqn.15}
F(r) = \frac{r - 4\pi r^3 [\varepsilon(r) - P(r)]}{r - 2M(r)},
\end{equation}
\begin{equation}\label{eqn.16}
\begin{aligned}
Q(r) = & \frac{4\pi r [5\varepsilon(r) + 9P(r) + \frac{\varepsilon(r) + P(r)}{\partial P(r)/\partial \varepsilon(r)}]}{r - 2M(r)} \\
& - 4 \left[ \frac{M(r) + 4\pi r^3 P(r)}{r^2 [1 - 2M(r)/r]} \right].
\end{aligned}
\end{equation}
In binary NS mergers, the tidal response is encoded in combined dimensionless tidal deformability parameter given by \cite{PhysRevLett.112.101101}
\begin{equation}\label{eqn.17}
\begin{aligned}
\tilde{\Lambda} & = \frac{16}{13} \frac{(M_1 + 12 M_2) M_{1}^4 \Lambda_1 + (M_2 + 12 M_1) M_{2}^4 \Lambda_2}{(M_1 + M_2)^5} .
\end{aligned}
\end{equation}
where $\Lambda_1$, $\Lambda_2$ are the dimesionless tidal deformabilities corresponding to stars with masses $M_1$ and $M_2$ respectively.

\subsection{Coupling parameters}

As mentioned earlier in Sec.-\ref{sec:intro}, in the present work, we implement GM1 parametrization \cite{1991PhRvL..67.2414G} with density-dependence of isovector-vector $\rho$-meson for meson-baryon couplings \cite{2017PhDT.......135S} and DD-MEX  parametrization \cite{TANINAH2020135065}. In case of GM1 parametrization, the density dependent coupling constant for the isovector $\rho$-meson is given by
\begin{equation}\label{eqn.18}
g_{\rho N}(n)= g_{\rho N}(n_{0}) e^{-a_{\rho}(x-1)}
\end{equation}
where $x=n/n_0$, while the coupling constants for $\sigma$, $\omega$-mesons are considered to be density-independent. 

In case of density-dependent model, the isoscalar meson-nucleon couplings are defined as
\begin{equation}\label{eqn.dd_isoscalar}
g_{i N}(n)= g_{i N}(n_{0}) f_i(x) \quad \quad \text{for }i=\sigma,\omega
\end{equation}
where, the function is given by
\begin{equation}\label{eqn.func}
f_i(x)= a_i \frac{1+b_i (x+d_i)^2}{1+c_i (x +d_i)^2}
\end{equation}
and the density dependence of isovector-vector $\rho$-meson coupling is given by eqn.-\eqref{eqn.18}.
Table-\ref{tab:1} provides the parameter values of GM1 and DD-MEX coupling parametrizations in nucleonic sector. In the standard GM1 parametrization, $g_{\rho N}$ is density independent and for the standard DD-MEX parametrization the coefficient $a_\rho$ is given in table-\ref{tab:1}. For variation in $L_{\text{sym}}$, it is evaluated by calibrating the coefficient $a_\rho$ without altering the other nuclear saturation properties. Table-\ref{tab:1} provides the values of $g_{i N}$ and $g_{\rho N}$ at $n_0$. Since the non-strange baryons do not couple with $\phi$-meson, $g_{\phi N}=g_{\phi \Delta}=0$.

The bare masses of baryons are considered as $m_N=939$ MeV, $m_\Lambda=1115.68$ MeV, $m_{\Sigma^+}=1189.37$ MeV, $m_{\Sigma^0}=1192.64$ MeV, $m_{\Sigma^-}=1197.45$ MeV, $m_{\Xi^-}=1321.71$ MeV, $m_{\Xi^0}=1314.86$ MeV and $m_{\Delta}=1232$ MeV. 
The saturation property parameters viz. $n_0$, saturation energy ($E_0$), incompressibility ($K_0$), $E_{sym}$, $L_{sym}$, curvature of symmetry energy ($K_{sym}$) and effective nucleonic Dirac mass ($m^*_N$) corresponding to these parametrizations are given in table-\ref{tab:2}. 
\begin{table} [h!]
\centering
\caption{The nuclear properties of the GM1 (A) and DD-MEX (B) CDF models at $n_0$.}
\begin{tabular}{ccccccc}
\hline \hline
 $n_0$ & $E_0$ & $K_0$ & $E_{sym}$ & $L_{sym}$ & $K_{sym}$ & $m^*_N/m_N$ \\
 (fm$^{-3}$) & (MeV) & (MeV) & (MeV) & (MeV) & (MeV) & \\
\hline
(A) 0.153 & $-16.30$ & 300 & 32.50 & 93.86 & 17.91 & 0.700 \\
(B) 0.152 & $-16.14$ & 267 & 32.27 & 49.58 & $-71.47$ & 0.556 \\
\hline
\end{tabular}
\label{tab:2}
\end{table}

The values of coefficient $a_{\rho}$ adjusted to estimate different values of $L_{\text{sym}}$ at $n_0$ for GM1 and DD-MEX parametrizations are provided in table-\ref{tab:3},
\begin{table} [h!]
\centering
\caption{$a_{\rho}$ coefficient values for various estimations of $L_{\text{sym}}(n_0)$ and corresponding $K_{\text{sym}}(n_0)$ for GM1 and DD-MEX coupling models.}
\begin{tabular}{c|cc|cc}
\hline \hline
 $L_{\text{sym}}(n_0)$ (MeV) & \multicolumn{2}{c|}{$a_\rho$} & \multicolumn{2}{c}{$K_{\text{sym}}(n_0)$ (MeV)} \\
\cline{2-5}
  & GM1 & DD-MEX & GM1 & DD-MEX \\
 \hline
 35 & 0.5893 & 0.8052 & $-127.13$ & $-34.44$ \\
 50 & 0.4390 & 0.6148 & $-129.67$ & $-72.21$ \\
 65 & 0.2888 & 0.4242 & $-105.17$ & $-75.72$ \\
 85 & 0.0885 & 0.1702 & $-30.43$ & $-27.06$ \\
\hline
\end{tabular}
\label{tab:3}
\end{table}
where the slope and curvature of $E_{\text{sym}}$ at $n_0$ are respectively given by
\begin{equation} \label{eqn.20}
\begin{aligned}
L_{\text{sym}}(n_0) & = 3 n_0 \left[\frac{\partial E_{\text{sym}} (n)}{\partial n} \right]_{n=n_0}, \\
K_{\text{sym}}(n_0) & = 9 n_0^2 \left[\frac{\partial^2 E_{\text{sym}} (n)}{\partial n^2} \right]_{n=n_0}.
\end{aligned} 
\end{equation}
with the nuclear symmetry energy defined as
\begin{equation} \label{eqn.19}
E_{\text{sym}}=\frac{1}{2} \left[\frac{\partial^2 (\varepsilon/n)}{\partial \alpha^2} \right]_{\alpha=0}
\end{equation}
where $\alpha=(n_n - n_p)/n$ is the asymmetry parameter. 

For the hyperonic sector, the vector couplings are implemented according to SU(6) symmetry and quark counting rule \cite{SCHAFFNER199435}. For the scalar couplings, we consider the optical potential values $U_{\Lambda}^N(n_0)=-30$ MeV, $U_{\Sigma}^N(n_0)=+30$ MeV and $U_{\Xi}^N(n_0)=-14$ MeV \cite{2015ApJ...808....8G, RevModPhys.88.035004} in symmetric nuclear matter (SNM). 
Ref.-\cite{2021arXiv210400421F} recently reported an attractive optical potential for $\Xi$-hyperons in SNM corresponding to $U_{\Xi}^N (n_0)\geqslant -20$ MeV. Table-\ref{tab:4} provides the scalar meson-hyperon coupling values at $n_0$.
\begin{table}[h!]
\centering
\caption{Scalar meson-hyperon coupling constants, $R_{\sigma Y}=g_{\sigma Y}/g_{\sigma N}$ (normalized to meson-nucleon coupling) for considered parametrizations in this work.}
\begin{tabular}{cccc}
\toprule
 & $\Lambda$ & $\Sigma$ & $\Xi$ \\
\hline
GM1 & 0.6164 & 0.4033 & 0.3047 \\
DD-MEX & 0.6172 & 0.4734 & 0.3088 \\
\hline
\end{tabular}
\label{tab:4}
\end{table}

For the $\Delta$-resonance sector, we consider the meson-$\Delta$ couplings as parameters. 
This is due to scarcity of $\Delta$-nucleon interaction experimental data. Experimental studies \cite{KOCH1985765, WEHRBERGER1989797, PhysRevC.81.035502} based on pion-nucleus scattering, $\Delta$-quartet excitations have reported to constrain meson-$\Delta$ resonance couplings. 
Recent studies \cite{Drago_PRC_2014, Kolomeitsev_NPA_2017} have narrowed the $\Delta$-potential ($V_{\Delta}$) in nuclear medium to be $-30~\text{MeV} + V_N \leq V_{\Delta} \leq V_N$ and $0 \leq R_{\sigma \Delta}-R_{\omega \Delta} \leq 0.2$ with $R_{\sigma \Delta}=g_{\sigma \Delta}/g_{\sigma N}$, $R_{\omega \Delta}=g_{\omega \Delta}/g_{\omega N}$.
Several works \cite{Li_PLB_2018, Ribes_2019, Cai_PRC_2015, 2021PhLB..81436070R, 2021MNRAS.507.2991T} have considered the vector couplings in the range values $R_{\omega \Delta} \in [0.6-1.2]$ and $R_{\rho \Delta} \in [0.5-3.0]$.
In the present discussion, we consider $R_{\omega \Delta}=1.10$ and $R_{\rho \Delta}=1.00$ in the vector coupling sector.
And in the scalar meson-$\Delta$ resonance coupling sector, we consider $R_{\sigma \Delta}=1.10, 1.20$.

\section{Results and Discussion}\label{sec:results}

\begin{figure} [h!]
  \begin{center}
\includegraphics[width=8.5cm,keepaspectratio ]{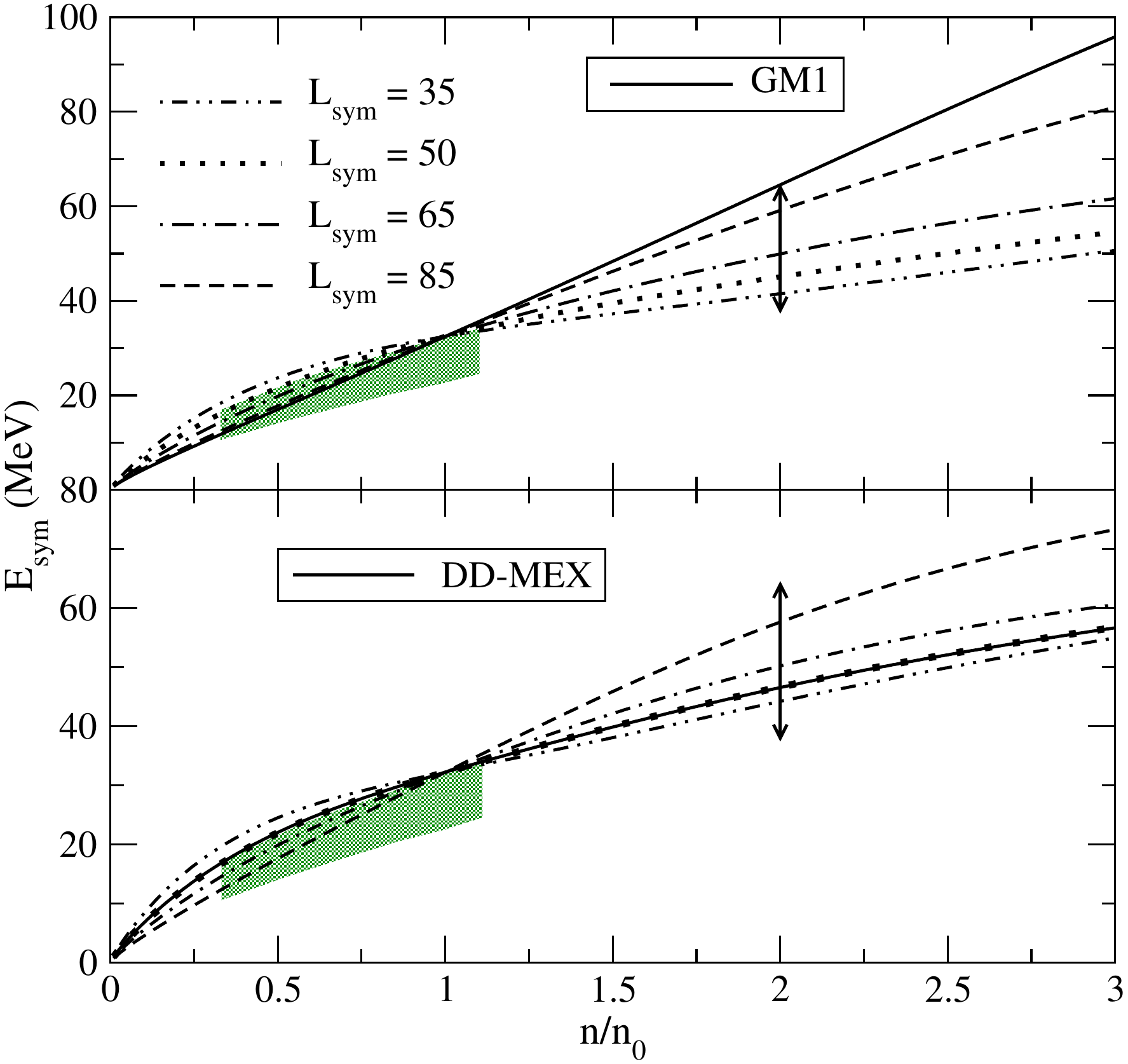}
\caption{Nuclear symmetry energy as a function of baryon number density (in units of $n_0$) for, upper panel: GM1 and lower panel: DD-MEX coupling parametrizations. The shaded regions denote the constraints on density-dependent symmetry energy from heavy-ion collision data \cite{2009PhRvL.102l2701T, TSANG2011400}. 
The constraint $38 \leqslant E_{\text{sym}}(2n_0)/\text{MeV} \leqslant 64$ \cite{2021Univ....7..182L} at $68\%$ confidence level obtained via analyses of data from recent NS observables and heavy-ion collisions is denoted by the vertical error bars.
The solid lines in both the panels represent the original coupling parametrizations. The other cases with adjusted values of $L_{\text{sym}}$ at $n_0$ are denoted by dot-dot-dashed ($L_{\text{sym}}=35$), dotted ($L_{\text{sym}}=50$), dash-dotted ($L_{\text{sym}}=65$) and dashed ($L_{\text{sym}}=85$) curves respectively.}
\label{fig:2}
\end{center}
\end{figure}

In this section, we report the numerical results for purely nucleonic (N), hypernuclear (NY) and $\Delta$-admixed hypernuclear (NY$\Delta$) matter compositions and investigate the effects of symmetry energy on dense matter EOS. 
In order to do so, as mentioned in Sec. \ref{sec:intro}, we implement the density-dependent modification in isovector $g_{\rho b}$ couplings within the framework of non-linear GM1 parametrization and consider the density-dependent coupling schemes with DD-MEX parametrization. We proceed by studying the effect of variation of $L_\text{sym}$ on different properties of matter and star.

\begin{figure} [h!]
  \begin{center}
\includegraphics[width=8.5cm,keepaspectratio ]{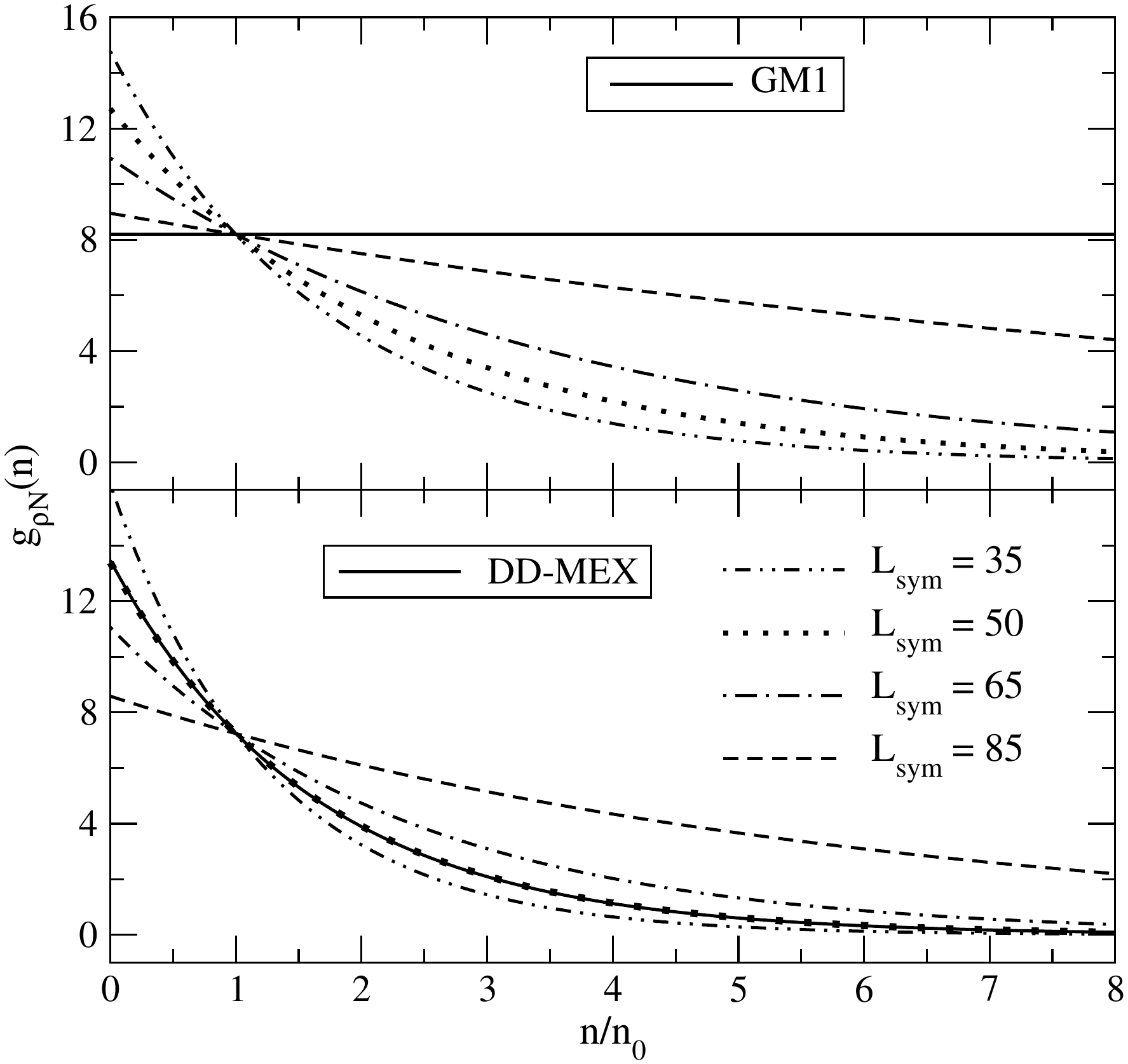}
\caption{Isovector coupling to nucleons as a function of baryon number density (in units of $n_0$) in case of GM1 (upper panel) and DD-MEX (lower panel) parametrizations. The different curves represent the same cases as captioned in fig.-\ref{fig:2}.}
\label{fig:1}
\end{center}
\end{figure}

\begin{figure*} [t!]
  \begin{center}
\includegraphics[width=15.0cm,keepaspectratio ]{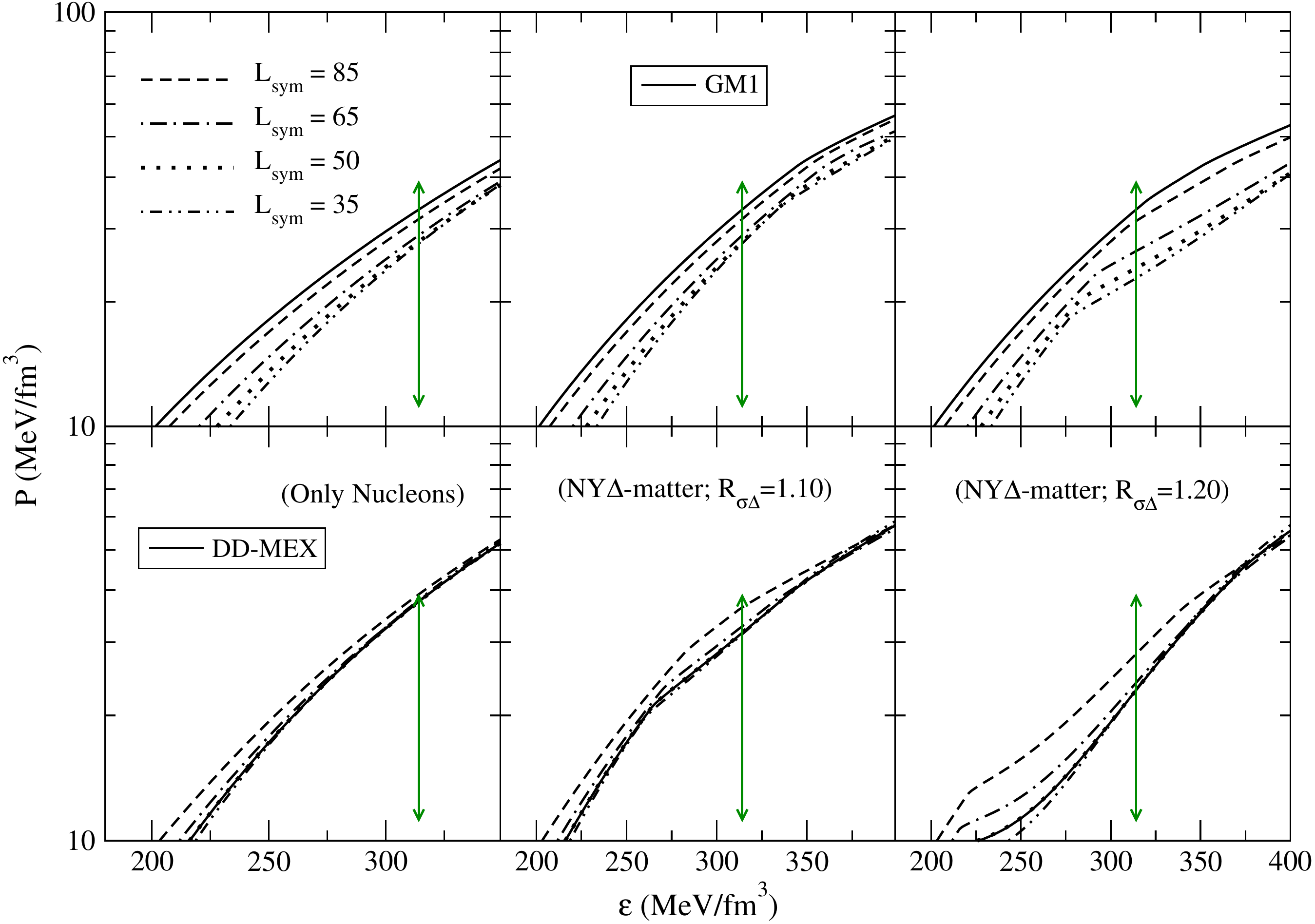}
\caption{Pressure variation as a function of energy density (EOS) for $T=0$ case with matter compositions as, left panels: pure nucleonic, middle panels: NY$\Delta$ ($R_{\sigma \Delta}=1.10$) and right panels: NY$\Delta$ ($R_{\sigma \Delta}=1.20$) for different $L_{\text{sym}}(n_0)$ values in upper panels: GM1 and lower panels: DD-MEX parametrizations. The different curves represent the same cases as captioned in fig.-\ref{fig:2}. The matter pressure constraint (vertical line) at $n\sim 2n_0$ is deduced from GW170817 \cite{LIGO_Virgo2018a} event data.}
\label{fig:3}
\end{center}
\end{figure*}
The behaviour of nuclear symmetry energy with varying baryon number density is plotted in fig.-\ref{fig:2} for different values of $L_\text{sym}$. In density regime $n<n_0$, cases with higher values of $L_{\text{sym}}$ yield lower values of $E_{\text{sym}}$ while the vice-versa is observed in case of higher density regimes ($n>n_0$). 
The experimental constraints on $E_{\text{sym}}(n)$ at sub-saturation densities shown by the shaded region in the fig.-\ref{fig:2} allows EOSs with $L_{\text{sym}}(n_0)\geqslant 50$ MeV. The values of $E_{\text{sym}}$ at $n_0$ are same for all values of $L_\text{sym}$ as it is constrained by the isovector coupling value at nuclear saturation.
This result is consistent with that of ref.-\cite{2021PhRvC.104a5802W} found considering NL3 \cite{1997PhRvC..55..540L} parametrization.
The constraint on $E_{\text{sym}}(2n_0)$ is broader and allows for almost all EOSs corresponding to $L_{\text{sym}}(n_0)$ values considered in this work.

Fig.-\ref{fig:1} displays the density-dependent nature of isovector couplings in variation with baryon number density for different values of $L_\text{sym}$ in both coupling schemes. In sub-saturation densities, it is observed that with lower $L_{\text{sym}}$ values, the isovector coupling values are larger. This behaviour is vice-versa in supra-saturation density regimes. At the saturation density, $g_{\rho N}$ values are identical owing to eqn.-\eqref{eqn.18}. With higher values of $L_{\text{sym}}$, the variation of $g_{\rho N}$ with baryon number density is found to be more steep. The $g_{\rho N}(n)$ coupling values with lower $L_{\text{sym}}$ approach zero at high density regimes resulting in similar  corresponding $E_{\text{sym}}(n)$ values at those densities.

The EOSs for different NS matter compositions (N, NY$\Delta$) are presented in fig.-\ref{fig:3} for GM1 parametrization in upper panels and for DD-MEX parametrization in lower panels.
The EOSs with modified isovector couplings within non-linear GM1 model as well as with DD-MEX parametrization are observed to lie well within bounds of the matter pressure constraint from GW170817 event data \cite{LIGO_Virgo2018a} shown by the vertical arrows in fig.-\ref{fig:3}. This is true for all the matter composition cases.
The prominent differences in EOSs are observed at low density regimes ($n \leqslant 0.4$ fm$^{-3}$).

\begin{table} [h!]
\centering
\caption{Threshold densities, $n_u$ (in units of $n_0$) for hyperons and $\Delta$-quartet in NY and NY$\Delta$ matter with varying $L_{\text{sym}}(n_0)$ values in GM1 parametrization.}
\begin{tabular}{c|c|cc|cc}
\hline \hline
 Model & \multicolumn{1}{c|}{$R_{\sigma \Delta}=0$} & \multicolumn{2}{c|}{$R_{\sigma \Delta}=1.10$} & \multicolumn{2}{c}{$R_{\sigma \Delta}=1.20$} \\
\cline{2-6}
  & $n_u^{Y}$ ($n_0$) & $n_u^{Y}$ ($n_0$) & $n_u^{\Delta}$ ($n_0$) & $n_u^{Y}$ ($n_0$) & $n_u^{\Delta}$ ($n_0$) \\
 \hline
  GM1 & 2.25 & 2.25 & 2.89 & 2.29 & 2.11 \\
 $L_{\text{sym}}=35$ & 2.54 & 2.68 & 2.22 & 2.95 & 1.87 \\
 $L_{\text{sym}}=50$ & 2.49 & 2.57 & 2.27 & 2.84 & 1.90 \\
 $L_{\text{sym}}=65$ & 2.42 & 2.43 & 2.35 & 2.66 & 1.94 \\
 $L_{\text{sym}}=85$ & 2.30 & 2.30 & 2.61 & 2.39 & 2.04 \\
\hline
\end{tabular}
\label{tab:5}
\end{table}

\begin{figure*} %[h!]
  \begin{center}
\includegraphics[width=15cm,keepaspectratio ]{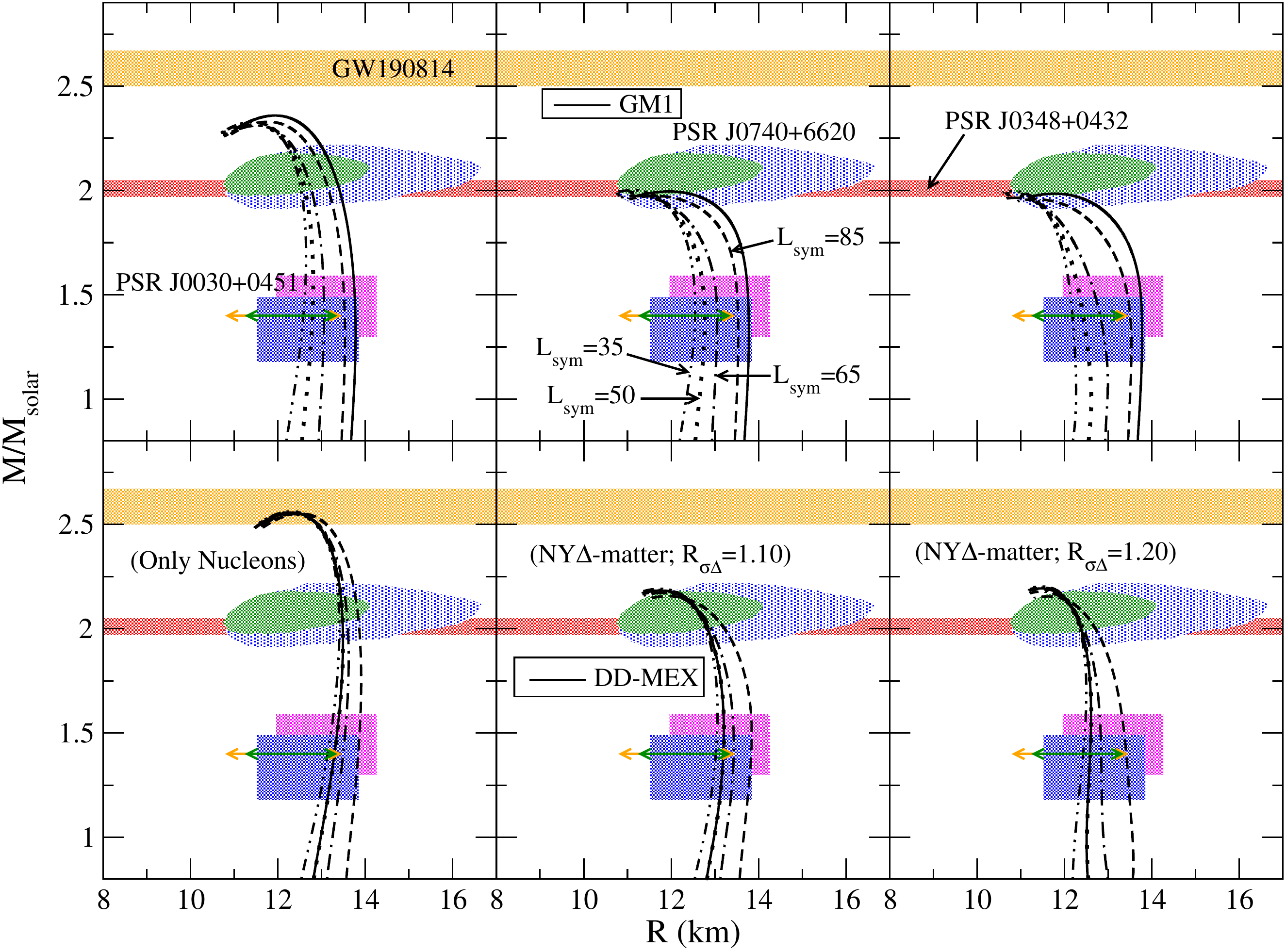}
\caption{Solutions of TOV equations corresponding to left panels: pure N-matter, middle panels: NY$\Delta$ ($R_{\sigma \Delta}=1.10$) and right panels: NY$\Delta$ ($R_{\sigma \Delta}=1.20$) EOSs displayed in fig.-\ref{fig:3} for upper panels: GM1 and lower panels: DD-MEX parametrizations. The different curves represent the same cases as captioned in fig.-\ref{fig:2}. The astrophysical observable constraints from GW190814 \cite{2020ApJ...896L..44A}, PSR J$0740+6620$ \cite{2021arXiv210506980R, 2021arXiv210506979M}, PSR J$0348+0432$ \cite{2013Sci...340..448A} and PSR J$0030+0451$ \cite{2019ApJ...887L..24M, 2019ApJ...887L..21R} are represented by shaded regions. The horizontal lines represent the joint radius constraints from PSR J$0030+0451$ and GW170817 event data for a typical $1.4~M_{\odot}$ NS \cite{Jiang_2020, 2020PhRvD.101l3007L}.}
\label{fig:4}
\end{center}
\end{figure*}
Tables-\ref{tab:5} and \ref{tab:6} provide the threshold densities of heavier baryons in NY and NY$\Delta$ matter with GM1 and DD-MEX coupling parametrizations respectively. It is observed that low values of $L_{\text{sym}}$ shifts the onset of hyperons to higher density regimes. While the opposite behaviour is seen in case of $\Delta$-resonances. Higher values of normalized scalar meson-$\Delta$ couplings denote attractive $\Delta$-potentials in SNM which result in early appearance of $\Delta$-quartet in NS matter. Onset of $\Delta$-resonances delay the appearance of hyperons in NS matter.
\begin{table} [h!]
\centering
\caption{Similar to table-\ref{tab:5} but with DD-MEX parametrization.}
\begin{tabular}{c|c|cc|cc}
\hline \hline
 Model & \multicolumn{1}{c|}{$R_{\sigma \Delta}=0$} & \multicolumn{2}{c|}{$R_{\sigma \Delta}=1.10$} & \multicolumn{2}{c}{$R_{\sigma \Delta}=1.20$} \\
\cline{2-6}
  & $n_u^{Y}$ ($n_0$) & $n_u^{Y}$ ($n_0$) & $n_u^{\Delta}$ ($n_0$) & $n_u^{Y}$ ($n_0$) & $n_u^{\Delta}$ ($n_0$) \\
 \hline
  DD-MEX & 2.13 & 2.27 & 1.79 & 2.47 & 1.46 \\
 $L_{\text{sym}}=35$ & 2.15 & 2.31 & 1.77 & 2.51 & 1.44 \\
 $L_{\text{sym}}=50$ & 2.13 & 2.27 & 1.79 & 2.47 & 1.46 \\
 $L_{\text{sym}}=65$ & 2.09 & 2.19 & 1.82 & 2.39 & 1.47 \\
 $L_{\text{sym}}=85$ & 2.03 & 2.07 & 1.89 & 2.24 & 1.50 \\
\hline
\end{tabular}
\label{tab:6}
\end{table}

It is observed that onset of heavier baryons softens the EOSs marked by change in slopes as shown in fig.-\ref{fig:3}. Now at high matter densities, $g_{\rho b}(n)$ coupling values tend to approach zero resulting in less contribution to EOSs from $\rho$-meson fields.
With increase in attractive $\Delta$-potential, the onset of $\Delta^-$ shifts towards lower density regimes as marked by the kinks in fig.-\ref{fig:3}.

The mass-radius relationships obtained by solving TOV equations for non-rotating, spherically symmetric stars corresponding to the EOSs for N, NY$\Delta$ ($R_{\sigma \Delta}=1.10,~1.20$) matter with GM1 and DD-MEX parametrizations 
are displayed in different panels of fig.-\ref{fig:4}. 
For the crust region, Baym-Pethick-Sutherland (BPS) \cite{1971ApJ...170..299B} EOS is implemented maintaining thermodynamic consistency with modelling crust-core transition following ref.-\cite{2016PhRvC..94c5804F}.
It can be observed that almost all EOSs (in N, NY$\Delta$-matter compositions) fit within the limits of recent astrophysical constraints.
However, the joint constraints on radius of a $1.4~M_{\odot}$ NS \cite{Jiang_2020, 2020PhRvD.101l3007L} are satisfied by EOS models with $L_{\text{sym}} \leq 65$ MeV. Incorporation of $\Delta$-quartet further softens the EOS at lower density in addition to high density regimes leading to NS configurations with smaller radii as evident from middle and right panels of fig.-\ref{fig:4}. The candidature of the secondary compact component involved in GW190814 as a NS is still not completely univocal \cite{2020PhRvD.102d1301S, LI2020135812}, hence the maximum mass constraint from this candidate is not so stringent.
The variation of symmetry energy slope has slight impact on maximum mass NS configurations owing to the similar values of $M_{\text{max}}$ (refer to table-\ref{tab:7}). The softening of EOSs due to inclusion of $\Delta$-resonances is more protruding in density-dependent scenario.
The effect of varying $L_{\text{sym}}$ is least for pure nucleonic matter and large for NY$\Delta$-matter (with more attractive $\Delta$-potential).
For DD-MEX coupling parametrization, the maximum mass NS configuration with purely nucleonic matter reaches $\sim 2.55 M_{\odot}$ satisfying the mass constraint from GW190814 event. This is consistent with the results in ref.-\cite{PhysRevC.103.055814}.

\begin{figure} [h!]
  \begin{center}
\includegraphics[width=8.5cm, keepaspectratio ]{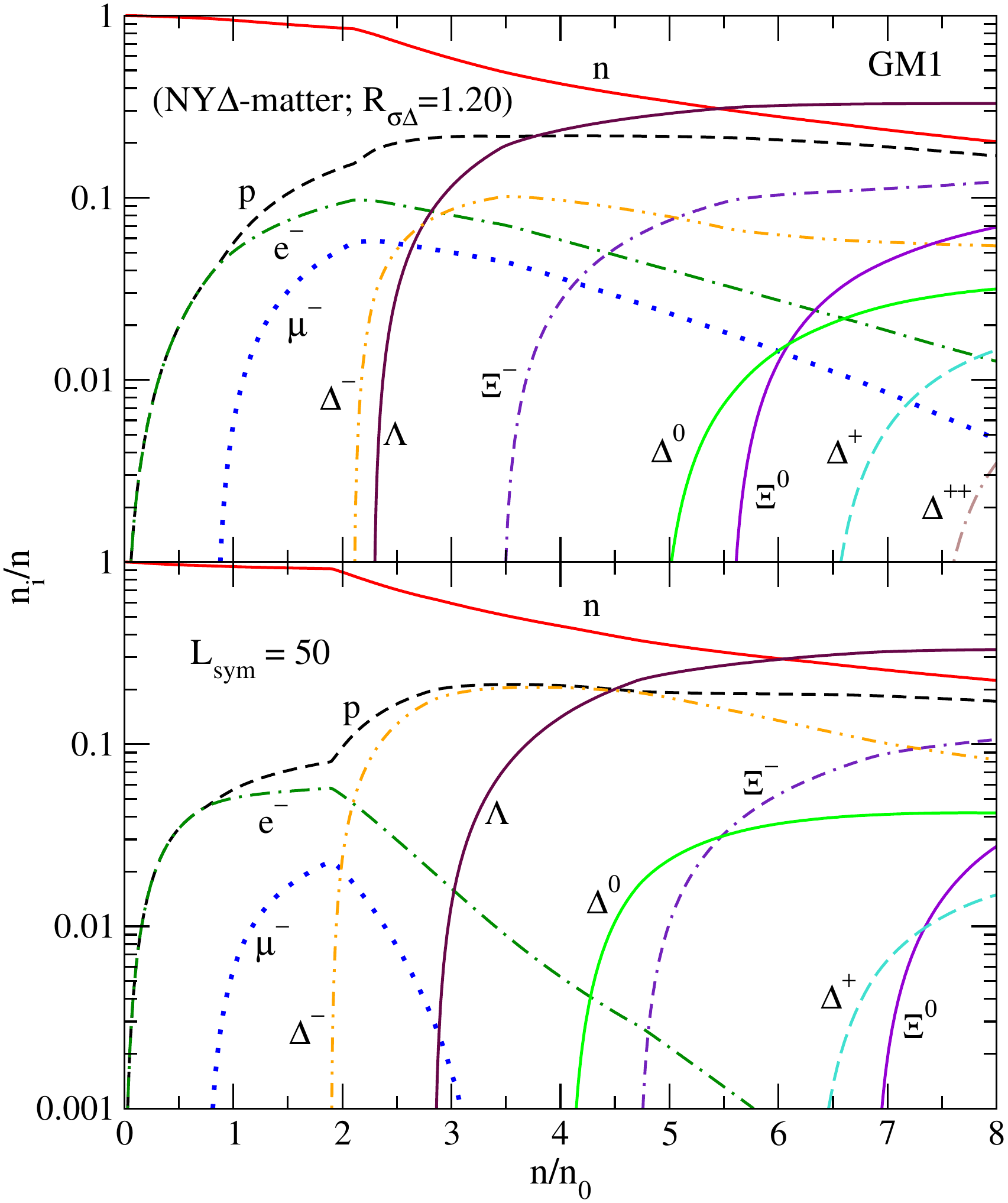}
\caption{Particle populations, $n_i$ (in units of $n$) as a function of baryon number density for NY$\Delta$-matter ($R_{\sigma \Delta}=1.20$) with upper panel: $L_{\text{sym}}(n_0)=93.86$ MeV (original), lower panel: $L_{\text{sym}}(n_0)=50$ MeV cases within non-linear coupling (GM1) parametrization.}
\label{fig:8}
\end{center}
\end{figure}

\begin{figure} [h!]
  \begin{center}
\includegraphics[width=8.5cm, keepaspectratio ]{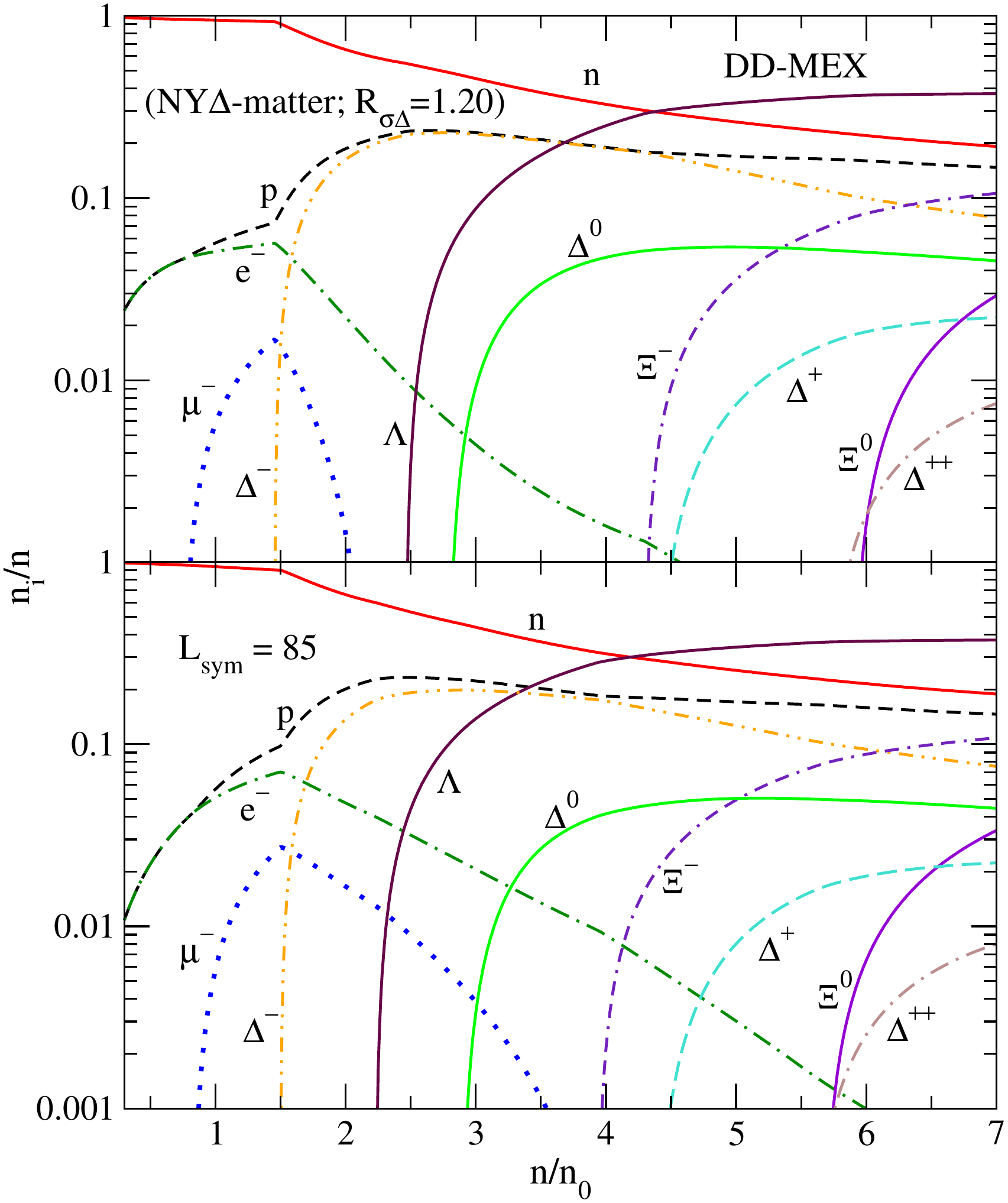}
\caption{Similar to fig.-\ref{fig:8} but with upper panel: $L_{\text{sym}}(n_0)=49.57$ MeV (original), lower panel: $L_{\text{sym}}(n_0)=85$ MeV cases within density-dependent coupling (DD-MEX) parametrization.}
\label{fig:9}
\end{center}
\end{figure}

The particle abundances in NY$\Delta$-matter composition ($R_{\sigma \Delta}=1.20$) with GM1 (without density dependent $g_{\rho N}$) and with $L_{\text{sym}}(n_0)=50$ MeV coupling parametrization models are shown in fig.-\ref{fig:8}.
It is observed that in case of density-dependent $g_{\rho N}(n)$ coupling (lower panel), the onset of hyperons is shifted to higher densities and early appearance of $\Delta$-resonances is favoured. This resulted in faster decrease of lepton populations in comparison to constant $g_{\rho N}(n_0)$ coupling case (upper panel).
In $g_{\rho N}(n_0)$ case, the $\Delta^-,~ \Xi^-, ~ e^-, ~ \mu^-$ composition provides the negative charge to balance the proton charge resulting in appearance of $\Lambda$-hyperons. At high density regimes, this negative particle composition leads to onset of $\Delta^0$ and $\Xi^0$ baryons. While in case of $g_{\rho N}(n)$ couplings, the onset of $\Delta^0$ baryons are due to charge neutrality condition maintained by $\Delta^-,~ e^-$ with protons.
Fig.-\ref{fig:9} displays the particle populations as a function of baryon number density similar to fig.-\ref{fig:8} but with DD-MEX parametrization for NY$\Delta$-matter.
In this too, it is observed that with decrasing value of $L_{\text{sym}}$, the onset of hyperons is shifted to higher densities while early onset of $\Delta$-quartet is favoured.
At sub-saturation densities, the charged particle abundances are enhanced with lower value of $L_{\text{sym}}$.
In case of $L_{\text{sym}}=49.57$ MeV (upper panel), the lepton populations are seen to decrease at a faster rate with rising matter density in comparison to $L_{\text{sym}}=85$ MeV (lower panel) case. This is because in the latter case $\Delta^-$ abundances fall short to maintain the charge neutrality condition with protons demanding lepton populations  to stay till higher density regimes. 

\begin{figure*}[!htb]
   \begin{minipage}{0.48\textwidth}
     \centering
     \includegraphics[width=8.5cm, keepaspectratio ]{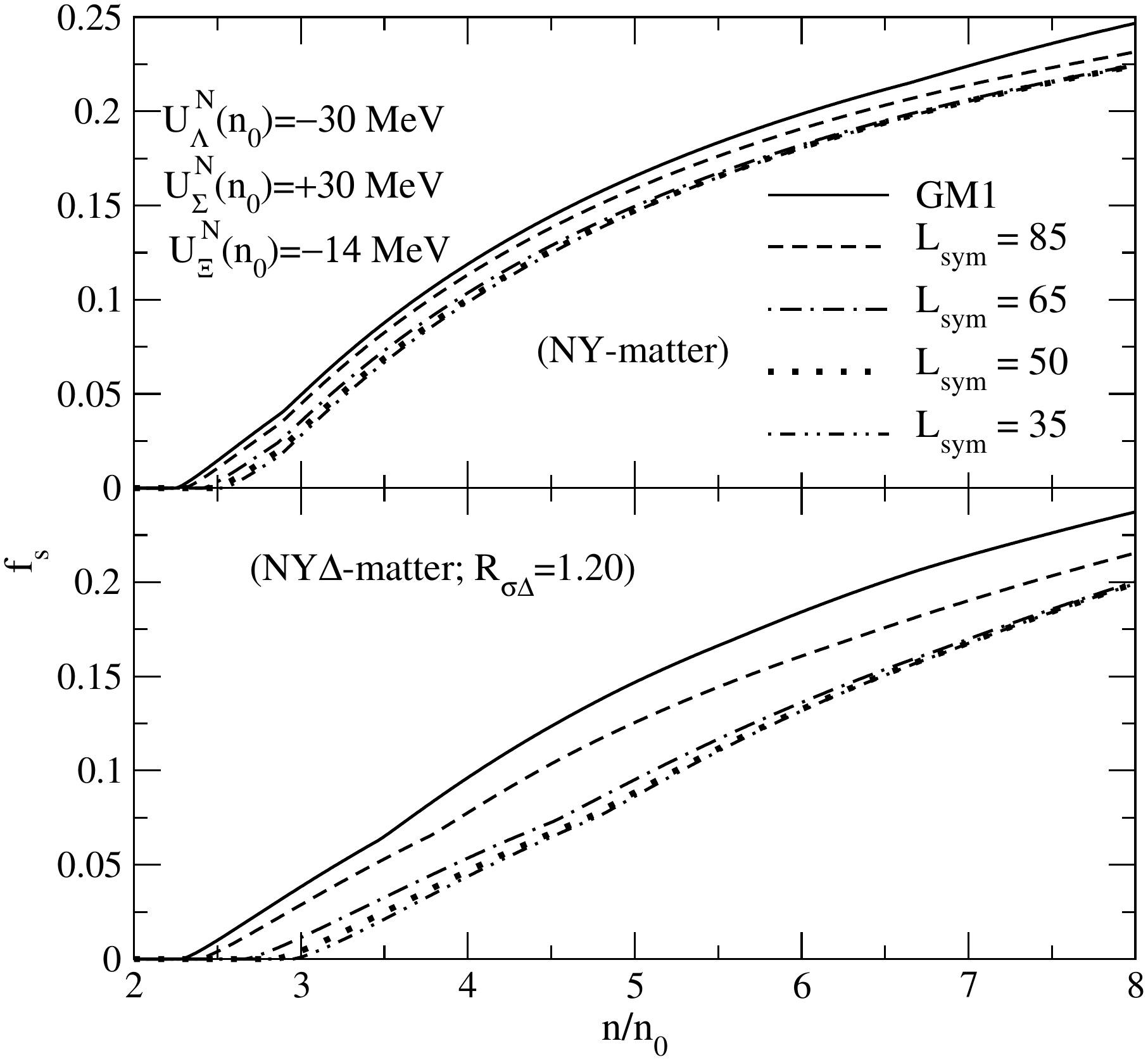}
   \end{minipage}\hfill
   \begin{minipage}{0.48\textwidth}
     \centering
     \includegraphics[width=8.5cm, keepaspectratio ]{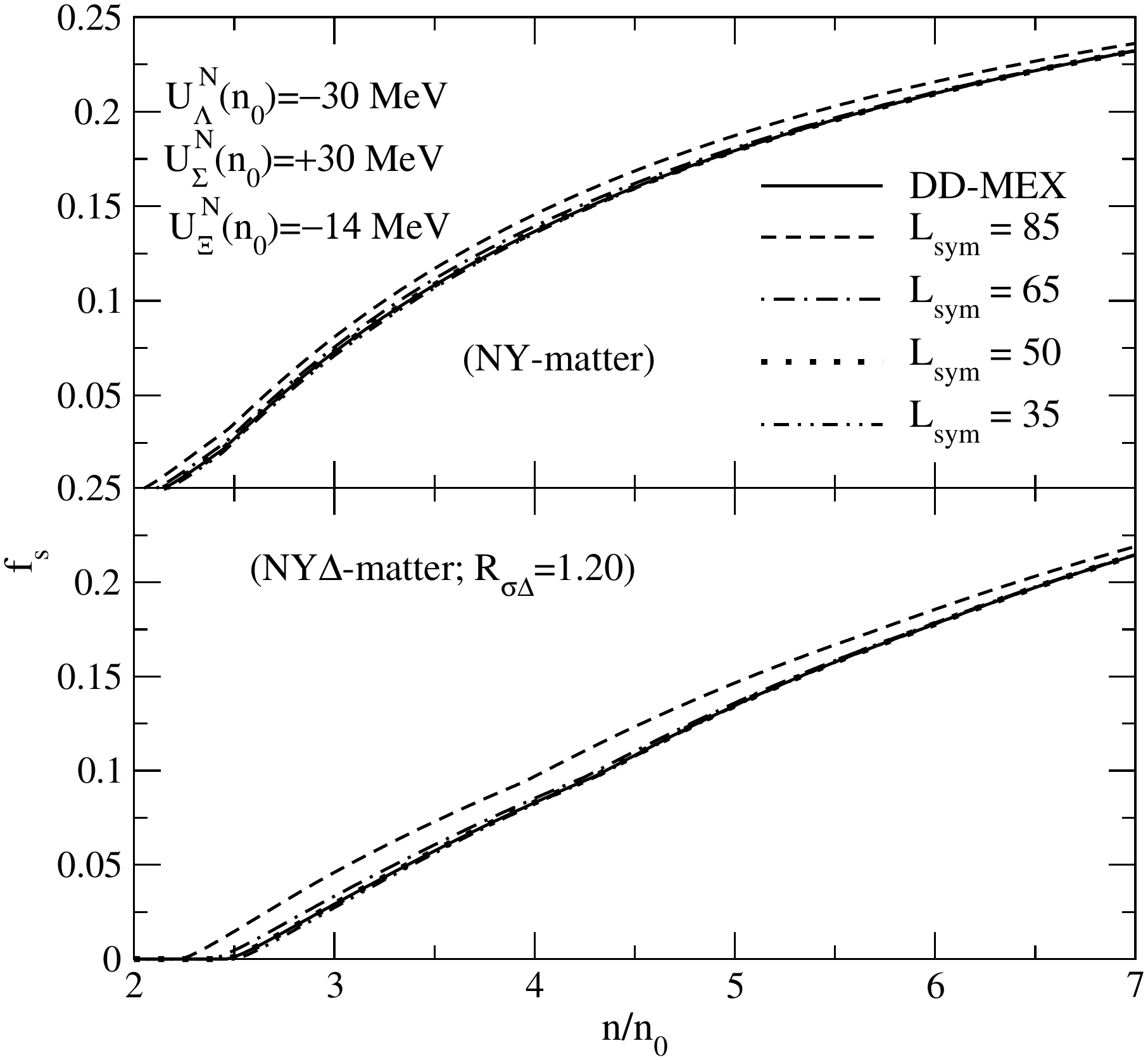}
   \end{minipage}
   \caption{Strangeness fraction, $f_s$ as a function of baryon number density for matter composition as, upper panels: NY and lower panels: NY$\Delta$ ($R_{\sigma \Delta}=1.20$) with varying $L_{\text{sym}}$ for left panels: GM1 and right panels: DD-MEX parametrizations. The different curves represent the same cases as captioned in fig.-\ref{fig:2}.} \label{fig:10}
\end{figure*}

\begin{figure} [h!]
  \begin{center}
\includegraphics[width=8.5cm, keepaspectratio ]{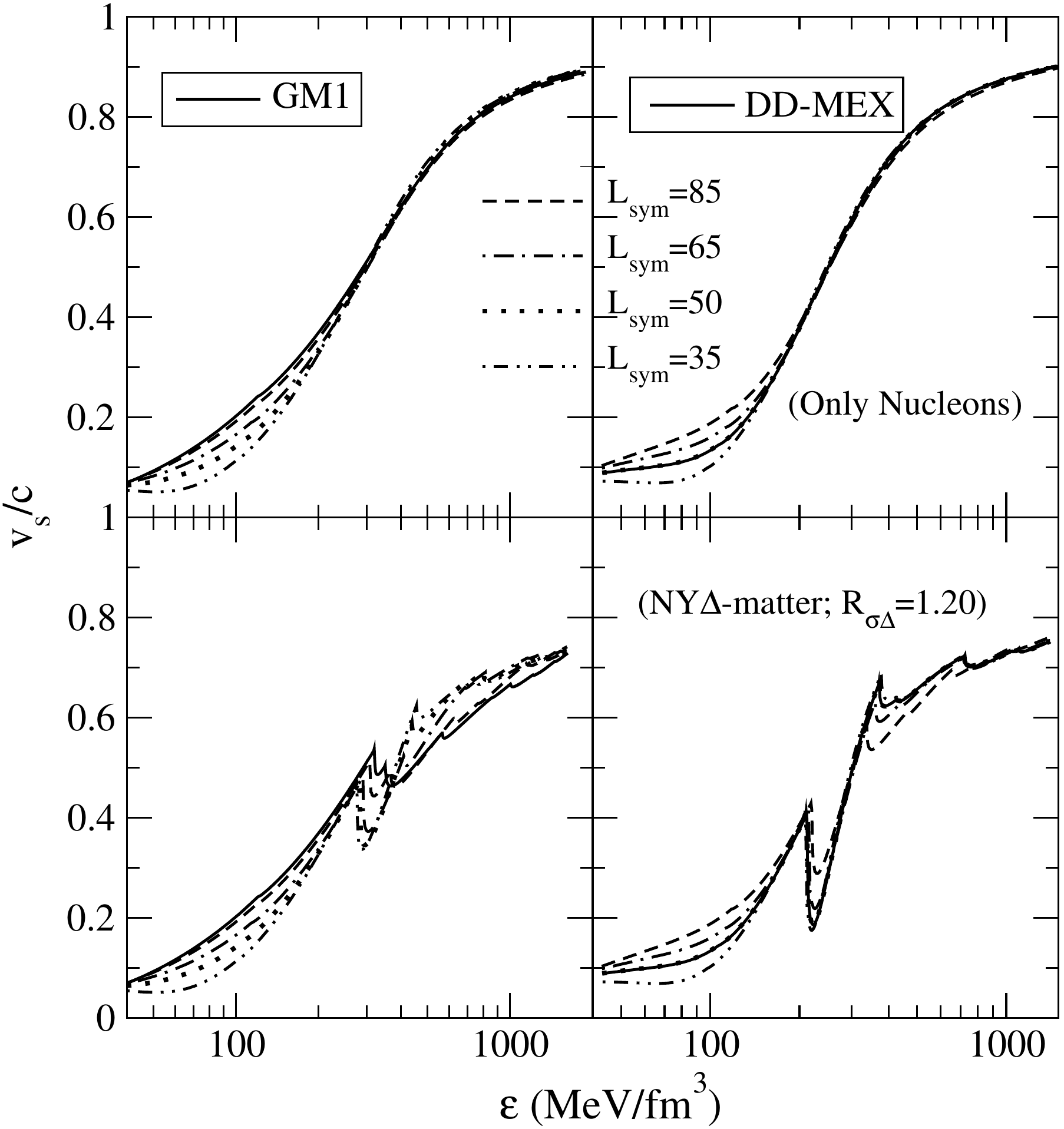}
\caption{Adiabatic sound velocity (in units of $c$) as a function of energy density for, upper panels: pure nucleonic, lower panels: $\Delta$-admixed hypernuclear matter ($R_{\sigma \Delta}=1.20$) with left panels: non-linear (GM1) and right panels: density-dependent (DD-MEX) coupling models. The different curves represent the same cases as captioned in fig.-\ref{fig:2}.}
\label{fig:13}
\end{center}
\end{figure}

In order to see the effect of varying $L_{\text{sym}}$ on hyperons we plot the strangeness fraction as a function of baryon number density in fig.-\ref{fig:10} which is defined as \cite{2011PhRvC..84f5810C},
\begin{equation}
f_s=\frac{1}{3} \frac{\sum_Y |s_Y|n_Y}{n}
\end{equation}
where $s_Y$, $n_Y$ denote the strangeness and number density of $Y$-th hyperon respectively.
It is seen that $f_s$ is sensitive to varying $L_{\text{sym}}$ and decreases with lowering of $L_{\text{sym}}$ values.
The shifting of hyperon threshold densities to higher densities with lower values of $L_{\text{sym}}$ (as seen in figs.-\ref{fig:8} and \ref{fig:9}) is also evident from fig.-\ref{fig:10}.
The similar strangeness fraction for $L_{\text{sym}}=65,50,35$ MeV cases at high densities relates with the almost similar values (approaching zero) of $g_{\rho N}(n)$ (refer to fig.-\ref{fig:1}). In both the coupling parametrization cases, similar $f_s$ values at high density regimes relates to the almost vanishing values of $\rho$-meson coupling.
The delaying appearance of hyperons into NS matter with onset of $\Delta$-quartet is also apparently seen in fig.-\ref{fig:10}.

Due to the utmost dependence of matter pressure explicitly over energy density in NS matter, it is noteworthy that the EOSs follow the causality condition (i.e. adiabatic speed velocity, $v_s$ to be subluminal) given by $v_s < c$.
Fig.-\ref{fig:13} displays the adiabatic speed of sound as a function of energy density for different matter compositions with variation in $L_{\text{sym}}$ values within non-linear and density-dependent CDF models.
It is observed that the EOSs considered in this work satisfy the causality condition. The effect of $L_{\text{sym}}$ is more prominent on the lower density regimes. This can be attributed to the diverse $g_{\rho N}$ coupling values at lower densities.
Lower values of $L_{\text{sym}}$ results in reduced $v_s$ at lower matter densities. Kinks in the lower panels denote onset of heavier baryons in NS matter.

\begin{table*} [t!]
\centering
\caption{NS properties evaluated from the EOS considering various matter compositions (N, NY, NY$\Delta$) with varying $L_{\text{sym}}(n_0)$ values with non-linear (GM1) and density-dependent (DD-MEX) coupling schemes. The maximum gravitational mass NS and its corresponding radius are denoted by $M_{\text{max}}$ (in units of $M_{\odot}$), $R$ (in units of km) respectively; central number density, central energy density, central matter pressure are represented by $n_c$ (in units of fm$^{-3}$), $\varepsilon_c$ and $P_c$ (in units of MeV/fm$^3$). Matter pressures at 2 and 6 times saturation densities are denoted by $P(2n_0)$, $P(6n_0)$ respectively. The global properties such as radius, compactness parameter, tidal Love number and tidal deformability for a $1.4M_\odot$ NS are given by $R_{1.4}$ (in units of km), $C_{1.4}$, $k_{2(1.4)}$ and $\Lambda_{1.4}$ respectively.}
\begin{tabular}{c|ccccccccccccc}
\hline \hline
Matter & \multicolumn{2}{c}{CDF Model} & $M_{\text{max}}$ & $R$ & $n_c$ & $\varepsilon_c$ & $P_c$ & $P (2n_0)$ & $P (6n_0)$ & $R_{1.4}$ & $C_{1.4}$ & $k_{2(1.4)}$ & $\Lambda_{1.4}$ \\
 composition & & & ($M_{\odot}$) & (km) & ($\text{fm}^{-3}$) & (MeV/fm$^3$) & (MeV/fm$^3$) & (MeV/fm$^3$) & (MeV/fm$^3$) & (km) &  & & \\
 \hline
 & & GM1 & 2.36 & 11.93 & 0.865 & 1116.75 & 500.67 & 30.48 & 574.12 & 13.77 & 0.150 & 0.100 & 882 \\
 & & $L_{\text{sym}}=85$ & 2.33 & 11.77 & 0.888 & 1145.36 & 512.52 & 28.62 & 553.83 & 13.53 & 0.153 & 0.098 & 785 \\
 & NL & $L_{\text{sym}}=65$ & 2.31 & 11.51 & 0.917 & 1185.47 & 547.87 & 25.54 & 549.66 & 13.06 & 0.158 & 0.096 & 640 \\
 & & $L_{\text{sym}}=50$ & 2.32 & 11.45 & 0.919 & 1186.87 & 555.04 & 24.22 & 554.15 & 12.79 & 0.162 & 0.096 & 581 \\
Pure & & $L_{\text{sym}}=35$ & 2.33 & 11.42 & 0.916 & 1180.98 & 553.01 & 23.59 & 556.13 & 12.58 & 0.164 & 0.102 & 568 \\
 \cline{2-14}
Nucleonic & & DD-MEX & 2.56 & 12.33 & 0.776 & 1000.40 & 487.24 & 32.26 & 704.96 & 13.29 & 0.156 & 0.106 & 773 \\
Matter & & $L_{\text{sym}}=85$ & 2.55 & 12.50 & 0.767 & 988.90 & 469.57 & 34.70 & 698.51 & 13.84 & 0.149 & 0.105 & 939 \\
 & DD & $L_{\text{sym}}=65$ & 2.55 & 12.36 & 0.777 & 1003.21 & 486.80 & 32.74 & 703.14 & 13.49 & 0.153 & 0.104 & 821 \\
 & & $L_{\text{sym}}=50$ & 2.56 & 12.33 & 0.776 & 1000.68 & 487.42 & 32.27 & 704.93 & 13.30 & 0.155 & 0.105 & 772 \\
 & & $L_{\text{sym}}=35$ & 2.56 & 12.31 & 0.773 & 995.26 & 484.09 & 32.31 & 705.36 & 13.14 & 0.157 & 0.108 & 748 \\
 \hline
 & & GM1 & 1.99 & 11.97 & 0.926 & 1126.57 & 317.96 & 30.48 & 312.09 & 13.77 & 0.150 & 0.101 & 882 \\
 & & $L_{\text{sym}}=85$ & 1.98 & 11.72 & 0.964 & 1179.58 & 345.34 & 28.62 & 310.77 & 13.53 & 0.153 & 0.098 & 785 \\
 & NL & $L_{\text{sym}}=65$ & 1.98 & 11.41 & 1.001 & 1233.99 & 382.86 & 25.54 & 317.59 & 13.06 & 0.158 & 0.095 & 639 \\
 & & $L_{\text{sym}}=50$ & 2.00 & 11.37 & 0.994 & 1222.21 & 379.64 & 24.22 & 320.41 & 12.79 & 0.162 & 0.096 & 581 \\
Hypernuclear & & $L_{\text{sym}}=35$ & 2.01 & 11.36 & 0.983 & 1203.98 & 371.28 & 23.59 & 321.37 & 12.58 & 0.164 & 0.102 & 568 \\
 \cline{2-14}
Matter & & DD-MEX & 2.18 & 12.00 & 0.875 & 1082.30 & 362.19 & 32.26 & 395.89 & 13.29 & 0.156 & 0.106 & 777 \\
 & & $L_{\text{sym}}=85$ & 2.16 & 12.14 & 0.876 & 1085.95 & 357.77 & 34.70 & 390.54 & 13.83 & 0.149 & 0.105 & 937 \\
 & DD & $L_{\text{sym}}=65$ & 2.17 & 12.02 & 0.882 & 1093.24 & 367.13 & 32.74 & 395.07 & 13.49 & 0.153 & 0.104 & 821 \\
 & & $L_{\text{sym}}=50$ & 2.18 & 12.00 & 0.876 & 1082.86 & 362.48 & 32.27 & 395.88 & 13.30 & 0.155 & 0.105 & 774 \\
 & & $L_{\text{sym}}=35$ & 2.19 & 11.99 & 0.869 & 1071.27 & 356.43 & 32.31 & 396.04 & 13.14 & 0.157 & 0.108 & 748 \\
 \hline
 %%%%%%%%%%%%%%%%%%%%%%%%%%%%%%%%%%%%%%
 %%%%%%%%%%%%%%%%%%%%%%%%%%%%%%%%%%%%%%
 & & GM1 & 1.99 & 11.95 & 0.928 & 1130.21 & 320.17 & 30.48 & 312.53 & 13.77 & 0.150 & 0.101 & 882 \\
 & & $L_{\text{sym}}=85$ & 1.97 & 11.63 & 0.980 & 1204.54 & 360.38 & 28.62 & 312.22 & 11.63 & 0.153 & 0.098 & 785 \\
 & NL & $L_{\text{sym}}=65$ & 1.97 & 11.19 & 1.045 & 1302.99 & 428.33 & 25.54 & 324.02 & 13.05 & 0.158 & 0.095 & 637 \\
$\Delta$-admixed & & $L_{\text{sym}}=50$ & 1.99 & 11.13 & 1.042 & 1297.38 & 429.21 & 24.22 & 328.62 & 12.77 & 0.162 & 0.095 & 570 \\
Hypernuclear & & $L_{\text{sym}}=35$ & 2.00 & 11.11 & 1.031 & 1278.87 & 420.59 & 23.59 & 330.15 & 12.55 & 0.165 & 0.100 & 551 \\
 \cline{2-14}
Matter & & DD-MEX & 2.18 & 11.75 & 0.911 & 1138.96 & 405.08 & 28.03 & 406.19 & 13.19 & 0.157 & 0.102 & 717 \\
($R_{\sigma \Delta}=1.10$) & & $L_{\text{sym}}=85$ & 2.16 & 11.92 & 0.909 & 1138.40 & 395.67 & 33.15 & 398.31 & 13.83 & 0.149 & 0.104 & 935 \\
 & DD & $L_{\text{sym}}=65$ & 2.17 & 11.76 & 0.918 & 1152.14 & 411.28 & 29.43 & 405.19 & 13.43 & 0.154 & 0.102 & 788 \\
 & & $L_{\text{sym}}=50$ & 2.18 & 11.75 & 0.911 & 1139.52 & 405.38 & 28.06 & 406.18 & 13.20 & 0.157 & 0.102 & 720 \\
 & & $L_{\text{sym}}=35$ & 2.19 & 11.74 & 0.904 & 1127.36 & 398.93 & 27.51 & 406.38 & 13.02 & 0.159 & 0.104 & 686 \\
 %%%%%%%%%%%%%%%%%%%%%%%%%%%%%%%%%%%%
 %%%%%%%%%%%%%%%%%%%%%%%%%%%%%%%%%%%%
 \hline
 & & GM1 & 1.99 & 11.78 & 0.962 & 1179.86 & 345.22 & 30.48 & 313.94 & 13.77 & 0.150 & 0.101 & 881 \\
 & & $L_{\text{sym}}=85$ & 1.96 & 11.34 & 1.037 & 1290.93 & 410.47 & 28.62 & 316.04 & 13.50 & 0.153 & 0.097 & 773 \\
 & NL & $L_{\text{sym}}=65$ & 1.97 & 10.87 & 1.102 & 1392.46 & 488.48 & 24.65 & 332.45 & 12.82 & 0.161 & 0.089 & 541 \\
$\Delta$-admixed & & $L_{\text{sym}}=50$ & 1.98 & 10.82 & 1.094 & 1377.87 & 485.28 & 22.32 & 338.37 & 12.43 & 0.166 & 0.086 & 449 \\
Hypernuclear & & $L_{\text{sym}}=35$ & 2.00 & 10.81 & 1.083 & 1358.24 & 476.19 & 20.88 & 340.33 & 12.18 & 0.169 & 0.083 & 395 \\
 \cline{2-14}
Matter & & DD-MEX & 2.19 & 11.48 & 0.938 & 1177.66 & 439.80 & 18.67 & 416.19 & 12.59 & 0.164 & 0.090 & 505 \\
($R_{\sigma \Delta}=1.20$) & & $L_{\text{sym}}=85$ & 2.15 & 11.62 & 0.945 & 1193.37 & 436.81 & 24.49 & 405.46 & 13.41 & 0.154 & 0.094 & 720 \\
 & DD & $L_{\text{sym}}=65$ & 2.18 & 11.48 & 0.947 & 1194.49 & 447.69 & 20.08 & 414.88 & 12.84 & 0.161 & 0.089 & 565 \\
 & & $L_{\text{sym}}=50$ & 2.19 & 11.48 & 0.938 & 1177.94 & 439.93 & 18.69 & 416.17 & 12.59 & 0.164 & 0.090 & 506 \\
 & & $L_{\text{sym}}=35$ & 2.20 & 11.47 & 0.930 & 1165.22 & 433.33 & 18.30 & 416.43 & 12.43 & 0.166 & 0.093 & 487 \\
 \hline \hline
\end{tabular}
\label{tab:7}
\end{table*}

Now we move to examine the effect of $L_\text{sym}$ on the star properties.
The properties of NS for variation of $L_\text{sym}$ along with matter properties with different EOSs considered in this work have been displayed in table-\ref{tab:7}. $L_{\text{sym}}$ has practically no effect on the maximum mass of the star family. However, $L_{\text{sym}}$ variation has a commendable impact on the radius of NS configurations. This feature is already clear from the fig.-\ref{fig:4}. We have tabulated the comparative values of radius for typical $1.4~M_\odot$ mass NS ($R_{1.4}$). With increasing value of $L_{\text{sym}}$ the $R_{1.4}$ increases. Consequently, the compactness $C_{1.4}$ decreases. Similar impact is also observed in case of tidal deformability $\Lambda_{1.4}$ with increase of $L_{\text{sym}}$ softness decreases. 
A recent study \cite{2020Sci...370.1450D} based on the joint analysis GW170817 and GW190425 events data reported a radius bound $10.94\leq R_{1.4}/\text{km}\leq 12.61$ at $90\%$ confidence level.
From the EOSs considered in this work, it can be inferred that to satisfy the said $R_{1.4}$ range, the conditions of $\Delta$-resonances onset into NS matter composition and $L_{\text{sym}}(n_0)\leqslant 50$ MeV are favourable.
Following the $69 \leq L_{\text{sym}}(n_0)/\text{MeV} \leq 143$ range deduced from recent PREX-2 data, it is to be noted that DD-MEX parameterization satisfies the $\Lambda_{1.4}$ upper bound (GW170817 event) for $L_{\text{sym}}(n_0)\leq 85$ with $R_{\sigma \Delta}=1.20$.

\begin{figure} [h!]
  \begin{center}
\includegraphics[width=8.5cm, keepaspectratio ]{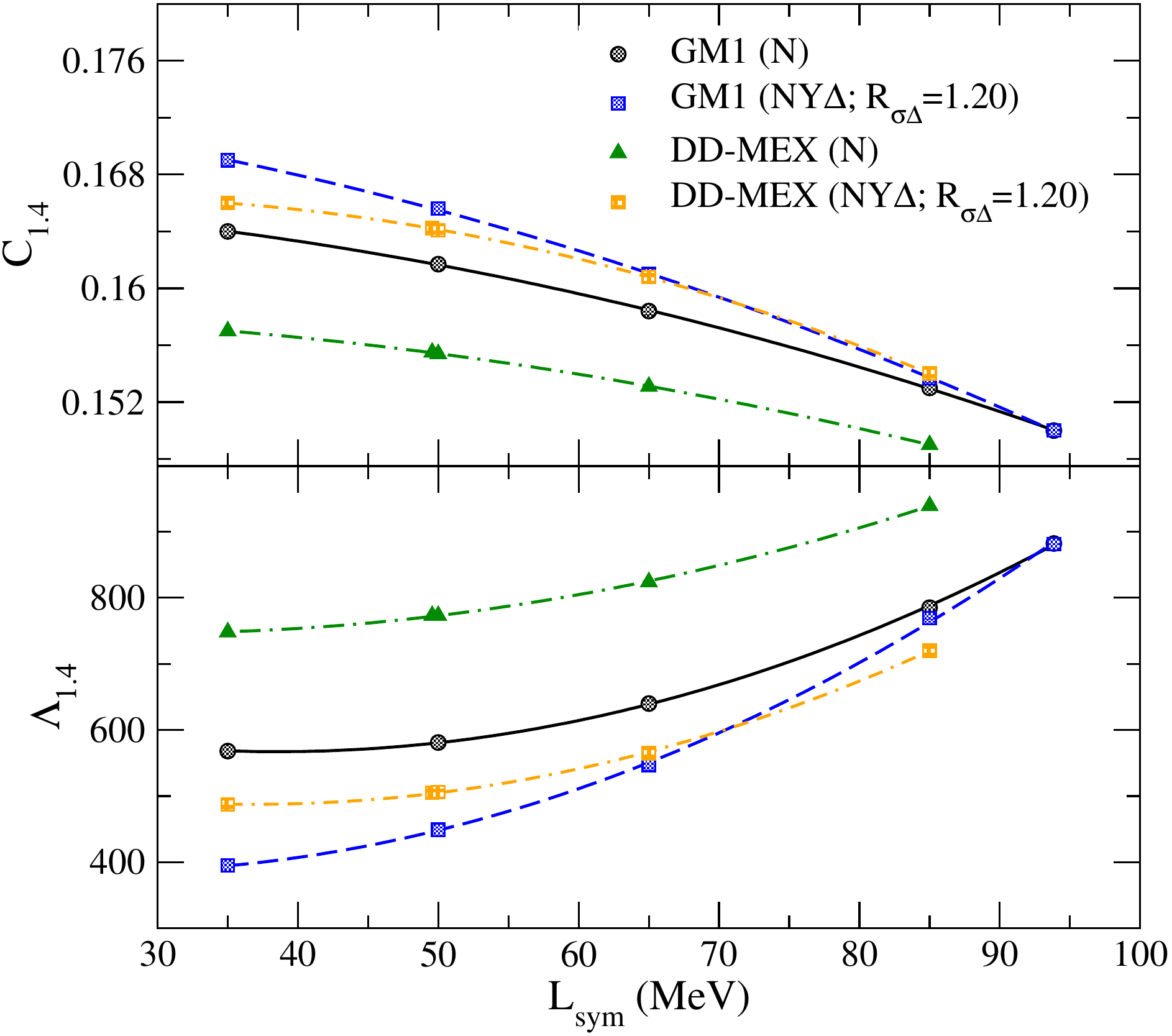}
\caption{Upper panel: Compactness parameter, lower panel: tidal deformability as a function of $L_{\text{sym}}$ considering matter composition to be pure N, NY$\Delta$ ($R_{\sigma \Delta}$=1.20) with non-linear (GM1) and density-dependent (DD-MEX) coupling models. The different curves denote the EOS models as labelled.}
\label{fig:16}
\end{center}
\end{figure}

The variations of compactness parameter and tidal deformability of $1.4~M_\odot$ NS with $L_{\text{sym}}$ considering various matter compositions are shown in fig.-\ref{fig:16}.
The softness decreases in both parametrizations following similar trend of convergence towards higher $L_{\text{sym}}$ values. This relates to the fact that lowering of $L_{\text{sym}}$ shifts the onset of $\Delta$-quartet to lower density regimes thus increasing compactness and decreasing tidal deformability.
The quadratic fit of compactness parameter and tidal deformability as a function of $L_{\text{sym}}$ for different matter compositions with GM1 and DD-MEX parametrizations is given by,
\begin{equation} \label{eqn.fit}
C_{1.4}~\text{or,}~\Lambda_{1.4} = a~L_{\text{sym}}^2 + b~L_{\text{sym}} + c,
\end{equation} 
where the coefficient $a$, $b$ and $c$ values are provided in table-\ref{tab:8}.

\begin{table} [h!]
\centering
\caption{Coefficient values of the quadratic fits in eqn.-\eqref{eqn.fit}. The coefficient of determination, $\mathcal{R}^2\sim 0.999$ for all the fits considered in this work.}
\begin{tabular}{c|c|ccc}
\hline \hline
 \multicolumn{2}{c|}{CDF Model} & \multicolumn{1}{c}{$a$} & \multicolumn{1}{c}{$b$} & \multicolumn{1}{c}{$c$} \\
\hline
   & GM1 (N) & $-1.83\times 10^{-6}$ & $-2.10\times 10^{-6}$ & 0.1663 \\
$C_{1.4}$  & GM1 (NY$\Delta$) & $-2.01\times 10^{-6}$ & $-6.51\times 10^{-5}$ & 0.1738 \\
  & DD-MEX (N) & $-1.53\times 10^{-6}$ & $2.38\times 10^{-5}$ & 0.1581 \\
  & DD-MEX (NY$\Delta$) & $-3.31\times 10^{-6}$ & 0.0002 & 0.1646 \\
\hline
   & GM1 (N) & 0.1023 & $-7.884$ & 719.10 \\
$\Lambda_{1.4}$  & GM1 (NY$\Delta$) & 0.1082 & $-5.617$ & 458.70 \\
  & DD-MEX (N) & 0.0623 & $-3.676$ & 800.80 \\
  & DD-MEX (NY$\Delta$) & 0.0998 & $-7.339$ & 622.40 \\
  \hline
\end{tabular}
\label{tab:8}
\end{table}

\begin{figure} [h!]
  \begin{center}
\includegraphics[width=8.5cm, keepaspectratio ]{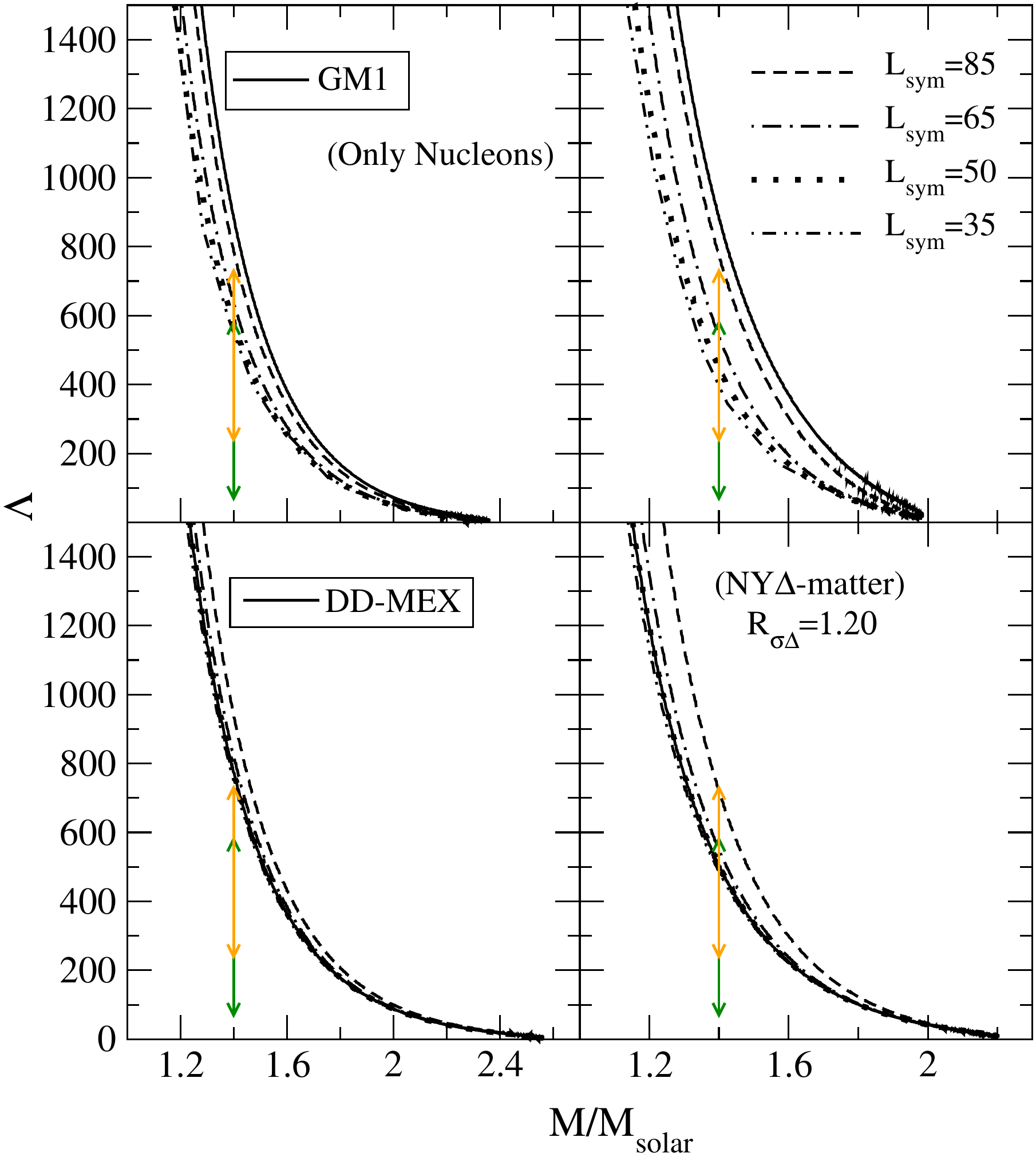}
\caption{Tidal deformability as a function of NS mass for, left panels: pure nucleonic, right panels: NY$\Delta$  ($R_{\sigma \Delta}=1.20$) matter with varying $L_{\text{sym}}$ values in upper panels: GM1, lower panels: DD-MEX parametrizations. The different curves represent the same cases as captioned in fig.-\ref{fig:2}. The vertical lines denote the bounds on $\Lambda_{1.4}$ deduced in refs.-\cite{LIGO_Virgo2018a, Jiang_2020}.}
\label{fig:12}
\end{center}
\end{figure}

Fig.-\ref{fig:12} shows the variation of dimensionless tidal deformability with NS mass corresponding to different values of density-dependent $L_{\text{sym}}$ parameter. As already mentioned, we observe that with higher values of $L_{\text{sym}}$, tidal deformability parameter value increases as the matter stiffens.
The effects of $L_{\text{sym}}$ is significant only in case of lower mass stars and for massive stars, these effects are inconsequential. In addition, the inclusion of heavier non-strange baryons softens the EOS at lower density regimes consequently decreasing $\Lambda$ or, assembling NS matter to be more compact. This relates with the results from refs.-\cite{Li2019ApJ,2021PhLB..81436070R}. 

\section{Summary and Conclusions}\label{sec:summary}

In this work, we discuss the density-dependent symmetry energy effects on dense matter EOSs with different matter compositions viz. pure nucleonic, hypernuclear, $\Delta$-admixed hypernuclear within CDF theory framework. We consider different values of symmetry energy slope $L_\text{sym}$ to introduce variation of symmetry energy with density.
The $L_{\text{sym}}$ at saturation is taken to be within the range of $35-85$ MeV.
The NS configurations evaluated from the EOSs considered in this work within this range of $L_\text{sym}$ satisfy the recent astrophysical observable constraints obtained from NICER (PSR J$0030+0451$ \cite{2019ApJ...887L..24M, 2019ApJ...887L..21R}, PSR J$0740+6620$ \cite{2021arXiv210506980R, 2021arXiv210506979M}) and GW \cite{LIGO_Virgo2017c, 2020ApJ...892L...3A} observations.

We find that with smaller values of $L_{\text{sym}}$, the EOS is evaluated to be softer around density range of $1-2~n_0$.
This is because of the corresponding lower values of $E_{\text{sym}}$ in the said density regimes.
This results in smaller radii alongside making the matter tidally less deformable for intermediate mass NSs viz. $1.4~M_\odot$. Although the range of $L_{\text{sym}}$ considered in this work is consistent with the astrophysical observations, the lower values of $L_\text{sym}$ are more favourable for the radius observations from NICER \cite{2019ApJ...887L..24M, 2019ApJ...887L..21R, 2021arXiv210506980R, 2021arXiv210506979M} as well as estimate of tidal deformability from GW observations.
While at the high density regimes, the EOS is similar for all $L_{\text{sym}}$ values. This attributes to the vanishing $\rho$ meson fields due to small (approaching zero) $g_{\rho N}$ coupling values.

Different values of $L_\text{sym}$ has also substantial effect on appearance of exotic component of matter. The lower values of $L_\text{sym}$ shift the threshold density for appearance of hyperons to higher side and favours early appearance of $\Delta$-quartet particles. The early appearance of $\Delta$ particles is also one of the causes for higher threshold density for appearance of hyperons with lower values of $L_{\text{sym}}$. The early appearance of $\Delta$ particles for lower values of $L_{\text{sym}}$ makes the EOS softer at lower density regime attributing to smaller radius for stars having mass $\sim 1.4~M_\odot$. Consequently, the compactness of the stars increases and the deformability decreases with the decreasing values of $L_{\text{sym}}$. The different values of $L_\text{sym}$ does not practically affect the maximum mass of the NS family as at higher density regime, the effect of different values of $L_{\text{sym}}$ on the EOS is negligible. This relates with the results from refs.-\cite{2011PhRvC..84f5810C, 2013PhRvC..87e5801P}. As for higher values of $L_{\text{sym}}$, the EOSs do not differ much, the values of compactness and tidal deformability merge at higher values of $L_{\text{sym}}$. Hence the possibility of exotic matter appearance with lower values of $L_{\text{sym}}$ is most favourable from all astrophysical observations.

However the recent update from nuclear physics sector (PREX-2) suggests higher values of $L_{\text{sym}}(n_0)$, which plays a tension with the astrophysical observables viz. tidal deformability, radius of a canonical $1.4~M_\odot$ NS. Considering the viability of inclusion of non-strange $\Delta$-baryons into NS dense matter EOS is seen to be a reasonable option.
Moreover, different choices of coupling models/parametrizations might or might not provide a feasible solution to this tension and so may have to be recalibrated.
Further analysis regarding this aspect is beyond the scope of this work and will be discussed in future studies.

The particle population of protons in $\beta$-equilibrated matter largely depends on the variation on nuclear symmetry energy with density. This in turn influences the threshold densities of nucleonic dUrca processes consequently in cooling processes. Several studies \cite{PhysRevC.74.045808, 2016PhRvC..94c5804F, 2018MNRAS.475.4347R, 2021PhLB..81436070R} have been done to understand the effects of $L_{\text{sym}}$ on this aspect. 
Such analysis with non-nucleonic matter composition following the recent update in isospin asymmetry parameter, $L_{\text{sym}}$ range will also be addressed in future works.

\begin{acknowledgements}
The authors thank the anonymous referee for the constructive comments which enhanced the quality of the manuscript.
\end{acknowledgements}

\bibliography{references}
\end{document}